\documentclass[lettersize,journal]{IEEEtran}

\usepackage{amsmath,amsfonts}
\usepackage{algorithmic}
\usepackage{array}
\usepackage[table]{xcolor}
\usepackage{colortbl}
\usepackage{booktabs}
\usepackage{tabularx}
\usepackage{multirow}
\usepackage{makecell}

\usepackage{graphicx}
\usepackage{adjustbox}
\usepackage[caption=false,font=normalsize,labelfont=sf,textfont=sf]{subfig}

\usepackage{textcomp}
\usepackage{url}
\usepackage{verbatim}
\usepackage{orcidlink}
\usepackage{balance}

\usepackage[most]{tcolorbox}
\usepackage{listings}
\usepackage{enumitem}
\usepackage{eso-pic}

\usepackage[table]{xcolor}   
\usepackage{colortbl}        
\usepackage{array}           
\usepackage{booktabs}        
\usepackage{tabularx}        
\usepackage{multirow}
\usepackage{makecell}
\usepackage{adjustbox}

\usepackage{tikz}
\usetikzlibrary{shapes.geometric, arrows.meta, positioning}
\usepackage[most]{tcolorbox}
\tcbuselibrary{listingsutf8}
\usepackage{listings}
\usepackage{fancyvrb}
\usepackage{framed}

\usepackage{enumitem}

\usepackage{fontawesome5}

\usepackage{hyperref}

\lstdefinestyle{jsonstyle}{
    basicstyle=\ttfamily\tiny\setstretch{0.8},
    keywordstyle=\color{blue},
    stringstyle=\color{red},
    numberstyle=\color{purple},
    commentstyle=\color{gray},
    breaklines=true,
    breakatwhitespace=false,
    showstringspaces=false,
    morestring=[b]",
    morecomment=[s]{/*}{*/},
    morecomment=[l]//,
    morekeywords={true,false,null},
    literate=
        *{:}{{{\color{blue}:}}}{1}
        {,}{{{\color{blue},}}}{1},
    aboveskip=5pt,
    belowskip=5pt,
    xleftmargin=0pt,
    xrightmargin=0pt,
}

\lstdefinelanguage{json}{
   morestring=[b]",
   morestring=[d]'
}

\usepackage{xargs}      
\usepackage{soul}       
\usepackage{xspace}     
\usepackage{xpunctuate} 
\usepackage{transparent}
\usepackage{tikz}
\usepackage{setspace}   
\usepackage{diagbox}
\usepackage{cleveref}
\usepackage{enumitem}

\definecolor{lightpink}{RGB}{237,157,202}
\definecolor{lightred}{RGB}{210,121,121}
\definecolor{lightorange}{RGB}{230,170,50}
\definecolor{lightgold}{RGB}{210,194,121}
\definecolor{lightgreen}{RGB}{121,210,121}
\definecolor{lightaqua}{RGB}{121,206,210}
\definecolor{lightblue}{RGB}{121,124,210}
\definecolor{lightpurple}{RGB}{153,102,255}
\definecolor{red}{RGB}{178,34,34}
\definecolor{gray}{RGB}{166,166,166}
\definecolor{peach}{RGB}{255,218,185}

\definecolor{Amber}{RGB}{255,190,11}
\definecolor{Pantone}{RGB}{251,86,7}
\definecolor{Rose}{RGB}{255,0,110}
\definecolor{Violet}{RGB}{131,56,236}
\definecolor{Azure}{RGB}{58,134,255}
\definecolor{mypurple}{RGB}{131,56,236}
\definecolor{myred}{RGB}{251,86,7}
\definecolor{mydarkred}{RGB}{252,102,32}
\definecolor{mydarkpurple}{RGB}{144,77,238}

\newcommandx{\yu}[2][1=] 
    {\setulcolor{orange}{\ul{#1}} \textcolor{orange}   
    {[\textbf{Yu:} #2]}}

\newcommandx{\jiawei}[2][1=] 
    {\setulcolor{lightgreen}{\ul{#1}} \textcolor{lightgreen}   
    {[\textbf{Jiawei:} #2]}}
    
\newcommandx{\munmun}[2][1=]{\setulcolor{teal}{\ul{#1}} \textcolor{teal}{[\textbf{Munmun:} #2]}}
\newcommandx{\john}[2][1=]{\setulcolor{blue}{\ul{#1}} \textcolor{blue}{[\textbf{John:} #2]}}

\usepackage{tcolorbox}
\tcbuselibrary{breakable}

\newtcolorbox{casebox}{
  breakable,
  colback=gray!5,
  colframe=gray!40,
  boxrule=0.5pt,
  arc=2pt,
  left=6pt,
  right=6pt,
  top=4pt,
  bottom=4pt,
  fonttitle=\bfseries,
}

\usepackage{tcolorbox}
\newtcolorbox{taxonomybg}{
  colback=gray!4,
  colframe=white,
  boxrule=0pt,
  arc=0pt,
  left=4pt,
  right=4pt,
  top=4pt,
  bottom=4pt,
  boxsep=0pt
}

\newcommand{\midlayer}[1]{%
\noindent\scalebox{1}{\faCaretRight}\ \textbf{#1}\ }

\def\BibTeX{{\rm B\kern-.05em{\sc i\kern-.025em b}\kern-.08em
  T\kern-.1667em\lower.7ex\hbox{E}\kern-.125emX}}

\newif\ifpreprint
\preprinttrue  

\AddToShipoutPictureFG*{%
  \ifpreprint
    \ifnum\value{page}=1
      \put(\LenToUnit{0.5\paperwidth-2in},\LenToUnit{\paperheight-0.4in}){%
        \parbox{4in}{\centering\sffamily\bfseries\footnotesize\color{gray}
          Preprint}%
      }
    \fi
  \fi
}

\begin{document}

\title{How Data Narratives Go Wrong: A Taxonomy of Issues Across the Data Communication Process}

\author{
Yu~Fu\,\orcidlink{0000-0001-5076-6299},
Jiawei~Zhou\,\orcidlink{0000-0003-2312-4359},
Sichen~Jin\,\orcidlink{0009-0006-3471-4716},
Munmun~De~Choudhury\,\orcidlink{0000-0002-8939-264X},
Cindy~Xiong~Bearfield\,\orcidlink{0000-0002-1451-4083},
and~John~Stasko\,\orcidlink{0000-0003-4129-7659}%
\thanks{Yu Fu is with the University of Central Florida, Orlando, FL, USA.\\
E-mail: yu.fu@ucf.edu}
\thanks{Jiawei Zhou, Sichen Jin, Munmun De Choudhury, Cindy Xiong Bearfield, and John Stasko are with Georgia Tech, Atlanta, GA, USA.}
}

\maketitle

\begin{abstract}
Data narratives increasingly shape public understanding, but their failures are rarely just isolated factual errors or deceptive charts. Instead, they emerge through a broader meaning-making process in which quantitative evidence is transformed into claims, representations, and arguments. While prior work has examined these failures across disparate fields (e.g., statistics, visualization, and fact-checking), the community lacks a holistic lens to explain how these issues arise, propagate, and compound. To address this gap, we introduce \textbf{TIC}, a \textbf{T}axonomy of \textbf{I}ssues in Data \textbf{C}ommunication, synthesized from prior literature and refined through the qualitative annotation of 700 real-world data narratives from fact-checking sites, research datasets, and controversial media. TIC organizes recurring breakdowns across six dimensions---data, analysis, visual encoding, text, reasoning, and interpretation---and situates them within a framework spanning analysis, narrative construction, and audience reception. Alongside the taxonomy and process framework, we contribute a qualitatively annotated case corpus with coding justifications and an interactive browsing interface. Collectively, these contributions provide a structured lens for diagnosing problematic data narratives and informing future sociotechnical support for trustworthy data communication.
\end{abstract}


\section{Introduction}

\IEEEPARstart{D}{ata} communication has become one of the primary lenses through which the public understands complex social, scientific, and political phenomena~\cite{Fu_2023_Morethan}. From tracking pandemic trends to interpreting climate risks and election results, quantitative evidence is routinely leveraged not only to inform but also to persuade~\cite{pandey_persuasive_2014}. However, the aura of objectivity surrounding ``the numbers''~\cite{porter_trust_1995} can obscure a fragile construction process in which narratives can break down not only through overt fabrication, but also through mechanisms such as misused statistics, cherry-picking, and systematic sampling biases.
This critique is not new: classic statistical skepticism---often summarized by the aphorism ``lies, damned lies, and statistics''---has long illustrated how valid numbers can be framed to support flawed conclusions~\cite{huff_how_1954}. More recently, influential critiques of data absence have highlighted how ``data silences'' can quietly shape public beliefs about risk and safety while maintaining an appearance of mathematical objectivity~\cite{criado_perez_invisible_2019, dignazio_klein_2020_datafeminism, Klein_DIgnazio_2024_DataFeminismAI}. Collectively, these examples underscore a critical reality: choices in how data are collected, analyzed, represented, and framed propagate through narratives to materially shape public understanding at scale.

At the core of data communication lies the \textit{data-driven narrative}, a communicative artifact that integrates quantitative evidence with interpretation through text and visual forms~\cite{segel_heer_2010}. Such narratives do more than present numbers; they organize evidence into interpretive accounts by situating data within context, foregrounding particular patterns, and guiding how audiences understand what is plausible, credible, or actionable. In practice, data-driven narratives encompass many forms, from textual claims that summarize key values and metrics~\cite{thorne_extensible_2017,fu_data_2024}, to visual artifacts such as tables and charts~\cite{few_show_2012}, to richly layered stories that integrate interactive visualizations with explanatory prose~\cite{segel_heer_2010,hullman_diakopoulos_2011}. Across this spectrum, \textit{text} and \textit{visualization} work in tandem~\cite{lan2026evolvingduetmodalitiessurvey}: visualizations surface patterns and relationships in data, while text articulates values, provides explanation, and steers interpretation. Seen this way, data-driven narratives are not merely finished artifacts, but products of a broader communication process in which evidence is transformed into claims, representations, and interpretations across actors and contexts.

Producing such data-driven narratives is a complex, iterative process of meaning-making that requires authors to move between data collection and analysis, interpretive framing, and multimodal integration~\cite{lee_more_2015, li_where_2024}. Visualization and HCI research have traditionally supported this authoring experience by developing systems that lower technical barriers or increase expressive power~\cite{Chen_2022_CrossData, Sultanum2021LeveragingVizflow, fu_dataweaver_2025}. However, comparatively less attention has been paid to supporting authors in constructing narratives that are not only expressive, but also evidentially grounded, appropriately contextualized, and logically warranted. This gap extends beyond authors: editors, fact-checkers, and general audiences also remain under-supported in their efforts to validate and interpret data narratives~\cite{Fu_2023_Morethan}. The challenge is further exacerbated by the rise of generative AI: while LLM-powered systems can accelerate narrative production~\cite{Sultanum2023DataTales:Articles, fu_dataweaver_2025, kantharaj_chart--text_2022}, they also risk introducing inaccuracies, omitting critical context, or overstating certainty, thereby compounding existing challenges in ensuring the trustworthiness and integrity of data-driven communication.

Across disciplines, researchers have pursued the shared goal of fostering more responsible data-driven communication, but their efforts remain largely siloed by the communication stage, modality, and disciplinary lens they examine. Statisticians have long critiqued analytical fallacies such as misleading aggregation and unwarranted causal inference~\cite{huff_how_1954, campbell2004flaws}. Visualization and HCI research have examined how design choices shape interpretation and can mislead perception~\cite{cairo_how_2019, mcnutt_2020, lo_misinformed_2022}. In parallel, work in NLP and data mining has focused on automated verification of numerical claims in textual forms~\cite{thorne_extensible_2017, wu_computational_2017}. These strands provide important insights, but they often treat statistical validity, visual representation, textual accuracy, and computational verification as separate problems. In practice, however, data-driven narratives rarely break down along such clean boundaries: text and visualization jointly frame evidence, support or obscure reasoning, and shape what audiences perceive as credible. Yet these cross-modal interactions, and their combined role in shaping credibility and interpretation~\cite{lundgard_accessible_2022, sultanum_instruction_2025}, remain insufficiently systematized.

Addressing this gap requires more than an inventory of misleading techniques; it demands an account that connects issue types to their locations within communication: \textit{what} breaks down in data-driven narratives, \textit{where} those breakdowns arise across the communication process, and \textit{how} this localization can inform sociotechnical support. To build this account, we combined a literature survey, qualitative annotation, and conceptual framework development. First, we synthesized 34 prior works to derive an initial categorization of documented issues. We then curated, analyzed, and annotated 700 real-world data-driven narratives drawn from fact-checking sites, prior research datasets, and controversial media sources, using these cases to refine and extend the categorization. Together, these steps produced \textbf{TIC}, a six-dimensional \textbf{T}axonomy of \textbf{I}ssues in data-driven \textbf{C}ommunication. Extending beyond visualization-centered taxonomies of misleading design, we situate TIC within Hall's encoding/decoding model~\cite{hall_encoding_1980} to locate recurring issues across the broader process through which quantitative evidence is constructed, represented, reasoned about, and interpreted. Our primary contributions are:
\begin{itemize}[leftmargin=*, topsep=0pt, itemsep=0pt] 
    \item \textbf{TIC taxonomy}: a six-dimensional, process-oriented taxonomy of recurring issues in data communication, spanning data, quantitative analysis, visual encoding, text and rhetoric, reasoning, and audience interpretation. TIC synthesizes prior work and surfaces additional patterns from our corpus.
    \item \textbf{Communication-process framework} that situates TIC across analysis, narrative construction, and audience interpretation, showing how breakdowns arise, propagate, and compound across actors, modalities, and contexts. It identifies leverage points for authoring, editorial review, fact-checking, and audience-facing support.
    \item \textbf{Annotated case corpus}, a curated collection of 700 real-world data narratives qualitatively annotated with TIC categories, coding justifications, metadata, and source links, accompanied by an interactive browsing interface. 
\end{itemize}

\section{Related Work}\label{section:rw}
Our work builds on prior research on problematic data communication, spanning statistical fallacies, misleading visualizations, and textual misrepresentation, as well as technical approaches for detecting and mitigating such problems.

\subsection{Problematic Data-driven Communication}\label{sec:problematic}

\subsubsection{Statistical Fallacies and Flaws}
Statistical misrepresentation has long been recognized as a byproduct of both intentional manipulation and methodological oversight. Classic work has documented how these issues arise from biased sampling, the misuse of averages, and misleading graphical summaries. Huff's \textit{How to Lie with Statistics}~\cite{huff_how_1954} introduced many of these issues to a broad audience, while Campbell catalogued statistical errors ranging from misapplied measures to flawed causal inference and misleading visual representation~\cite{campbell2004flaws}.
Strasak et al. showed that such pitfalls often accumulate across the research process, particularly in medical studies, underscoring the importance of rigorous methodology for preserving result integrity~\cite{Strasak2007}. Wainer further examined how improper scaling, reduced data density, and confusing chart designs impair comprehension~\cite{howard_wainer_how_1984}.

\subsubsection{Graphic Deception and Misleading Visualization} 
Beyond statistical reasoning, visualization research has examined the graphical dimension of data communication, producing a well-established body of work on how visual design choices bias or distort audience perception. Despite decades of design guidelines~\cite{cairo_how_2019, howard_wainer_how_1984, rogowitz_treinish_bryson_1996, Tufte01}, misleading visualizations remain pervasive across domains~\cite{vislies_gallary}. Pandey et al. systematically analyzed distortion techniques and demonstrated their measurable impact on perception~\cite{pandey_rall_satterthwaite_nov_bertini_2015}. Building on this line of work, McNutt et al. introduced the notion of ``visualization mirages,'' charts that appear credible but collapse under closer scrutiny of their underlying data or analytical process~\cite{mcnutt_2020}. Extending the scope, Lo et al. compiled a taxonomy of 74 visual issues from real-world misleading charts~\cite{lo_misinformed_2022}, while Lan et al. documented flawed practices surfaced in online design galleries~\cite{lan_i_2025}. Together, this literature has yielded fine-grained categorizations of misleading visual design practices, including axis and scale manipulations~\cite{correll_truncating_2020}, aspect-ratio distortions, and inappropriate chart types. More recent efforts have begun to empirically examine how design choices influence perception and decision-making~\cite{xiong_illusion_2019, holder_dispersion_2022}, as well as to investigate misleading facets beyond visual encodings, such as data validity and reasoning flaws~\cite{lisnic_misleading_2023, ge_calvi_2023, ge_v-framer_2024}.

\subsubsection{Textual and Rhetorical Misrepresentation}
Complementing visual encodings, textual elements play a critical role in shaping how data are interpreted. Text not only articulates specific values and describes graphical features but also steers attention, frames interpretation, and influences memory~\cite{borkin_beyond_2016}. In visualization research, text has traditionally been treated as supplementary---appearing in titles, captions, or annotations. Eye-tracking and user studies show that such elements strongly determine what viewers notice and recall~\cite{borkin_beyond_2016}. Misaligned or slanted titles can bias judgments and undermine credibility when narratives conflict with charts, and framing effects can override objective accuracy even when the visualizations themselves are correct~\cite{kong_frames_2018, Kong2019CHITrustAndRecall, obrien_testing_2018}. More recent work has begun to examine textual description at a finer granularity, showing how linguistic descriptors may drift from corresponding graphical features~\cite{bromley_what_2023,fu_data_2024}.


Prior work has largely examined isolated components of data-driven communication, producing rich but fragmented accounts of issues in statistical reasoning, visual design, and textual framing. TIC advances this body of work by integrating issues across data, analysis, visualization, text, reasoning, and audience interpretation, and by situating them within the broader communication process. This framing helps characterize not only what goes wrong, but also where issues arise and how they shape interpretation.

\subsection{Technological Support for Mitigation}
A complementary line of work has explored technological approaches for mitigating problematic data communication, ranging from automated verification to interactive tools.

\subsubsection{Automated Fact-checking}
Automated fact-checking has been extensively studied as a technical approach to assessing the credibility of public claims. Prior work typically decomposes fact-checking into subtasks such as claim detection, evidence retrieval, verdict prediction, and justification generation~\cite{Guo_2022_survey_automated_factchecking,nakov2021automated}. While much of this research focuses on textual claims, extensions have explored verification of data-driven statements through semantic parsing and query translation to structured data~\cite{thorne_extensible_2017,v_quantemp_2024}, as well as chart reasoning techniques that compare claims against visualized data~\cite{akhtar_chartcheck_2023}. Other systems translate natural language claims into executable queries, either automatically or in mixed-initiative settings; for example, AggChecker models uncertainty over candidate queries, while Scrutinizer incorporates human input to resolve ambiguity in query interpretation~\cite{jo2019verifying,Karagiannis_2020_Scrutinizer}. Collectively, these approaches advance automated and semi-automated verification capabilities, primarily targeting the correctness of individual claims or query results.

\subsubsection{Interactive Systems for Guardrails and Authoring}
Complementary work has focused on interactive systems that intervene earlier in the communication process, supporting human workflows while guarding against common errors and misleading practices. Linters and in-situ feedback tools~\cite{hopkins_visualint_2020, chen_vizlinter_2022} flag construction issues such as truncated axes, misleading scales, or inappropriate color use, while browser extensions and annotation systems surface potentially deceptive tactics during reading~\cite{Fan2022CHI}. Beyond error detection, alignment tools such as EmphasisChecker~\cite{kim_emphasischecker_2024} help synchronize textual framing with visual features, and SlopeSeeker~\cite{bendeck_slopeseeker_2024} operationalizes linguistic descriptors to assess correspondence between textual descriptions and visual trends. Other systems directly support the authoring of data-driven narratives by helping creators structure and refine intended stories~\cite{fu_dataweaver_2025,Sultanum2023DataTales:Articles}.

\medskip

To better guide such support, future systems need a structured account of the issues they are designed to surface, explain, or help users scrutinize. TIC provides this grounding by linking recurring communication issues to potential leverage points for computational and interactive support.
\section{Methodology}\label{method:taxonomy}
\begin{figure*}[tb]
    \centering
    \includegraphics[width=1\linewidth]{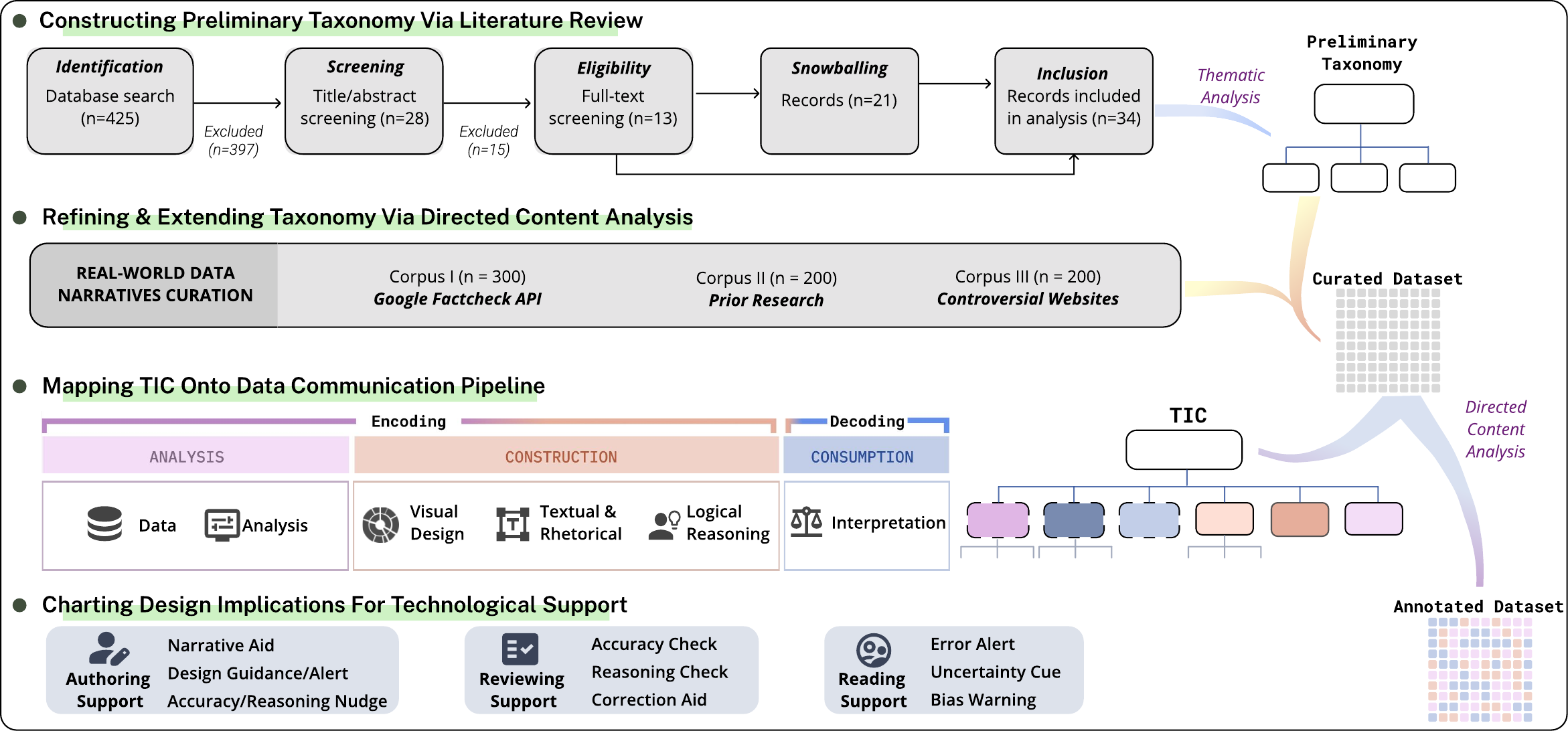}
    \caption{Overview of our methodology and contributions. We first construct a preliminary taxonomy of issues in data-driven communication through a systematic literature review and thematic synthesis. We then refine and extend this taxonomy via directed content analysis of curated real-world datasets (\autoref{tab:dataset_attributes}). Finally, we situate the taxonomy (\autoref{tab:taxonomy}) within the broader data communication process (\autoref{fig:framework}) to highlight potential leverage points for different stakeholders.
}
    \label{fig:method}
\end{figure*}

Our goal is to develop a taxonomy that systematically characterizes and organizes issues in the data-driven communication process. As shown in \autoref{fig:method}, we first construct a preliminary taxonomy through a structured literature review, and then refine and extend it by applying directed content analysis~\cite{hsieh2005three} to a diverse set of real-world data-driven narratives.

\subsection{Preliminary Taxonomy Construction}\label{sec:review}

\subsubsection{Literature Search}\label{sec:phase1}

To systematically identify prior work on problems in data-driven communication, we conducted a  PRISMA-guided literature review~\cite{page_prisma_2021} focusing on misleading representations, flawed reasoning, and misinterpretation of quantitative information. Drawing on the terminology used in these domains, we compiled a list of 97 keywords and phrases (provided in the supplementary materials) and used them to query the ACM Guide to Computing Literature, which indexes ACM venues and selected affiliated publishers (e.g., IEEE, Wiley, Springer, etc.), yielding 425 unique records. We then screened titles and abstracts for topical relevance, identifying 28 papers that proposed or applied categorization schemes related to problematic data-driven communication. Subsequent abstract review and full-text inspection retained 13 papers that aligned closely with our conceptual focus.  
To mitigate the limitations of keyword-based search, we then applied backward and forward snowballing~\cite{Wohlin2014Snowballing}, screening newly identified works using the same inclusion criteria and excluding papers that reiterated well-established issues without introducing new conceptual distinctions, resulting in 21 additional sources and a final corpus of 34 papers.

\subsubsection{Literature Synthesis}
The lead author reviewed and synthesized all eligible sources, with a second author independently verifying the inclusion set. For each work, we extracted issues documented in prior work on data-driven communication, resulting in a heterogeneous collection that varied in scope, granularity, and terminology. Because many of these issues were nested, overlapping, or articulated at different conceptual levels (e.g.,~\cite{lan_i_2025, lo_misinformed_2022}), we conducted an intermediate clustering step using thematic synthesis~\cite{thomas_harden_2008}. This process enabled us to group conceptually related issues, consolidate similar constructs, and identify higher-level themes that informed the structure of the preliminary taxonomy.

\subsection{Taxonomy Refinement and Extension}\label{sec:refine}

\subsubsection{Problematic Data Narratives Curation}
After synthesizing a preliminary taxonomy from prior literature, we next examined how well this taxonomy accounted for issues as they manifest in practice by applying it to curated real-world data-driven narratives. Existing fact-checking and misinformation datasets (e.g., LIAR~\cite{wang_2017_liar}, FEVER~\cite{thorne-etal-2018-fever}, MultiFC~\cite{Augenstein_2019_multiFC}, ClaimBuster~\cite{fatma_arslan_2020_3836810}) primarily focus on textual claims and largely overlook the quantitative and visual components that are central to many data narratives. Datasets on misleading visualizations~\cite{lo_misinformed_2022,lisnic_misleading_2023,lan_i_2025} often decouple graphics from their accompanying text or concentrate on narrow topical domains (e.g., COVID-19~\cite{lisnic_misleading_2023}). As a result, none of these resources alone adequately support refining and extending a taxonomy intended to characterize multimodal data-driven communication.
To address these limitations, we curated a dataset of \textit{real-world} data-driven narratives from diverse sources and communication contexts. This corpus provides the empirical foundation for refining the TIC taxonomy and investigating how these issues manifest in contemporary practice.

\begin{table*}[t]
\centering
\caption{Overview of curated real-world data-driven narratives by source, claim origin, topic, and modality.}
\small
\renewcommand{\arraystretch}{1.05}
\begin{tabular}{
  p{1.6cm}
  p{3.4cm}
  p{6.5cm}
  p{2.5cm}
  p{2cm}
}
\toprule
\textbf{Corpus (N)} & \textbf{Source} & \textbf{Primary Origin} & \textbf{Primary Topic} & \textbf{Modality} \\
\midrule
I (300) 
& Google Fact Check API
& Prominent individuals and viral posts
& Mixed
& Text \\

II (200) 
& Research datasets
& Social media and crowdsourced content
& COVID-19, Mixed
& Text + Chart \\

III (200) 
& Controversial websites
& Posts and articles published on these platforms
& Climate, Vaccines
& Mixed \\
\bottomrule
\end{tabular}
\label{tab:dataset_attributes}
\end{table*}

\textbf{Corpus I – Google Fact Check Tools}:  
We used the Google Fact Check Tools API~\cite{google_claim_search} to collect real-world, publicly circulated claims that had been evaluated by professional fact-checking organizations such as \textit{PolitiFact}~\cite{politifact_2024} and \textit{FactCheck.org}~\cite{factcheck.org_2024}. These claims originate from news articles, social media posts, and public speeches.
Because the API primarily indexes textual claims and does not explicitly distinguish data-driven content, we applied additional filtering to identify claims grounded in quantitative evidence. We selected domains where statistical claims are common (e.g., public health, economics, and sports) and manually curated topic-specific keywords (e.g., \textit{Cases, Deaths, Vaccinations} for COVID-19; \textit{Rate, Trends} for unemployment). This process yielded 756 candidate claims. After filtering for data-driven content and excluding paywalled items, we randomly sampled 300 claims rated as false or partially true for analysis.

\textbf{Corpus II – Datasets from Existing Research}:  
To complement our primary corpus with visualization-centered narratives, we incorporated datasets from prior research~\cite{lisnic_misleading_2023, akhtar_chartcheck_2023}. Specifically, from a dataset of COVID-19-related visualization posts on X.com annotated for potential errors~\cite{lisnic_misleading_2023}, we sampled 100 posts that included explicit narrative statements accompanying the visualizations. We also sampled 100 chart–annotation pairs from the ChartCheck dataset~\cite{akhtar_chartcheck_2023}, a benchmark collection of real-world charts paired with crowd-sourced claims and explanatory labels. For each sampled instance, we extracted both the textual narrative and its associated visualization for analysis.

\textbf{Corpus III – Controversial Sources}:  
To capture data-driven narratives with a higher likelihood of containing misleading or contentious claims, we curated an additional corpus from publicly accessible websites that regularly publish data-backed arguments challenging scientific or policy consensus. We began with Wikipedia's \textit{List of fake news websites}\footnote{\url{https://en.wikipedia.org/wiki/List_of_fake_news_websites}} as a starting point and employed a snowballing strategy to identify related sources referenced in discussions of controversial topics (e.g., climate change denial\footnote{\url{https://en.wikipedia.org/wiki/Climate_change_denial}}). From this process, we selected sites that consistently relied on statistical claims or visualizations to support their arguments, resulting in three representative sources: \textit{Watts Up With That}, \textit{CO\textsubscript{2} Coalition}, and \textit{Stop Mandatory Vaccination}\footnote{Sources: \url{https://wattsupwiththat.com}, \url{https://co2coalition.org/facts}, \url{https://www.stopmandatoryvaccination.com/}}. From each site, we manually extracted a diverse set of text-only and text–chart narratives, yielding 200 items for analysis.

\subsubsection{Directed Content Analysis}
Following directed content analysis, which combines existing theoretical frameworks with openness to emergent categories~\cite{hsieh2005three}, we refined the taxonomy through a structured three-phase qualitative annotation protocol. This process balanced deductive consistency with inductive flexibility. Annotations were conducted in a custom web interface that logged tagging decisions and justifications; details of the interface and annotation protocol are provided in the supplemental material. Four authors with complementary backgrounds (e.g., data journalism, visualization, misinformation) participated. In the \textit{pilot phase}, each author annotated 5\% of each corpus to become familiar with the interface, taxonomy, and coding process, followed by calibration meetings. In the \textit{open coding phase}, the lead author and a second annotator independently coded 15\% of the corpus, proposing new tags where needed; discrepancies and emergent tags were discussed in reconciliation meetings, yielding a consolidated codebook. In the \textit{main coding phase}, the lead author applied the refined codebook to the remaining data, documenting coding rationales while companion coders spot-checked examples to monitor consistency and support alignment.

\section{TIC: Taxonomy and Case Patterns}\label{sec:taxonomy}

\begin{figure*}[t]
    \centering
    \includegraphics[width=1\linewidth]{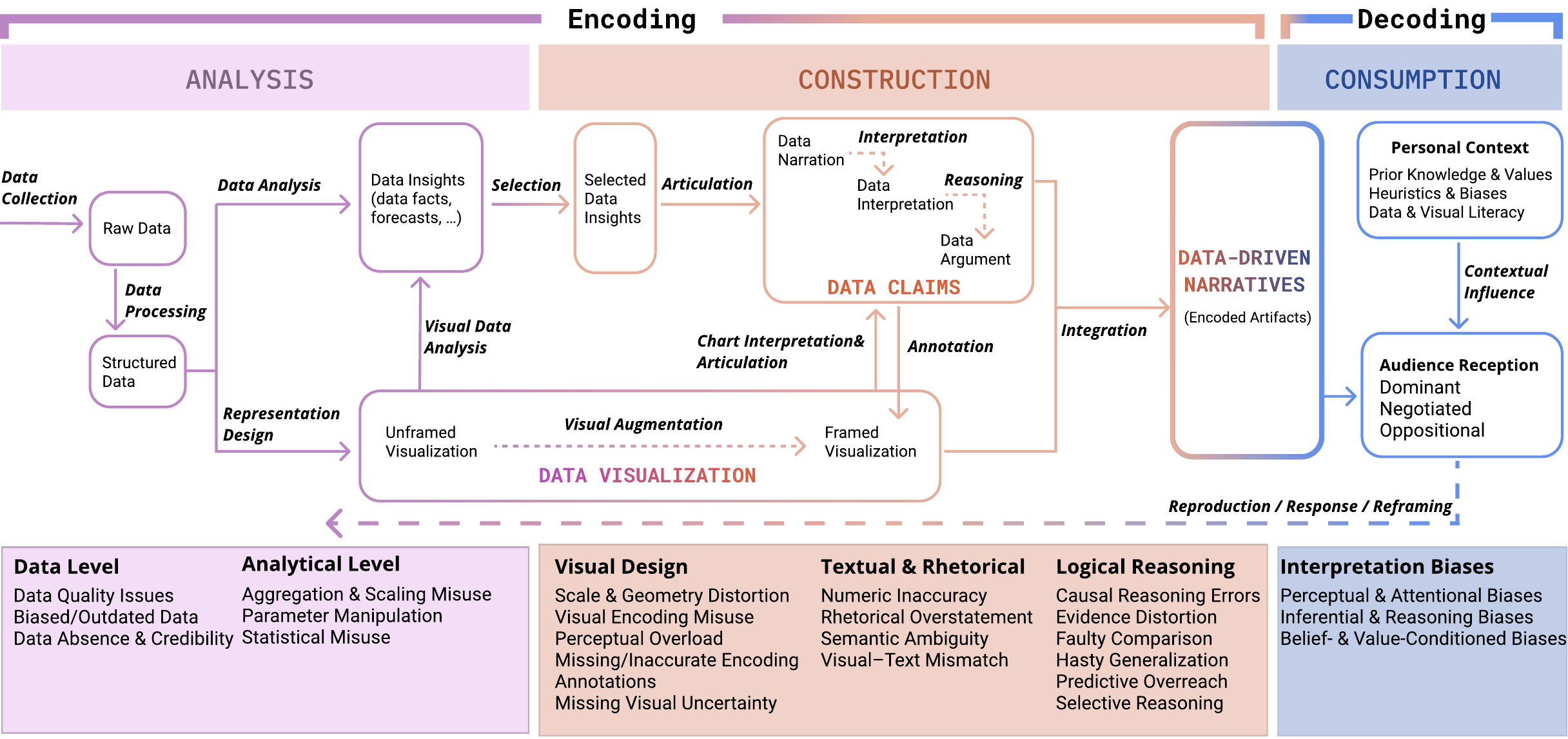}
    \caption{Framework situating the TIC taxonomy within the data communication process. 
    It maps six dimensions of issues onto activities of \textit{analysis}, \textit{construction}, and \textit{consumption}, 
    highlighting where problems arise, who is involved, and where leverage points for validation and support can be introduced.}
    \label{fig:framework}
\end{figure*}

This section introduces \textbf{TIC}, a multi-layer taxonomy of recurring issues in data-driven communication, synthesized from prior literature and analysis of real-world data narratives. TIC organizes these issues through a communication-process lens, distinguishing breakdowns associated with data, analysis, representation, reasoning, and interpretation. It is intended as an analytical lens rather than an exhaustive or mutually exclusive classification scheme: individual narratives may exhibit multiple interacting issues. We later situate these categories within the broader data communication process~(\autoref{fig:framework}); here, we define the taxonomy itself~(\autoref{tab:taxonomy}).

\renewcommand{\arraystretch}{1.1}
\newcommand{\SectionRow}[2]{%
  \multicolumn{2}{l}{\cellcolor{#1}\textbf{#2}}\\}

\begin{table*}[t]
\centering
\caption{TIC: A Taxonomy of Issues in Data-driven Communication}
\renewcommand{\baselinestretch}{1}\selectfont
\small
\begin{tabular}{p{6cm} p{11.3cm}}
{\small \textbf{Dimension $\rightarrow$ Type}} & {\small \textbf{Sub-Types (non-exhaustive)}}\\
\hline
\rowcolor{blue!15}\multicolumn{2}{l}{\textbf{1. Data Level Issues}}\\
Data Quality Issues &
Measurement Error; Missing/Incomplete Data \\[0em]

Biased/Outdated Data &
Sampling Bias; Outdated Data \\[0em]

Data Absence \& Credibility Issues &
No Supporting Data; Unreliable Data Source \\[0em]

\rowcolor{red!15}\multicolumn{2}{l}{{\textbf{2. Quantitative Analysis Issues}}}\\
Aggregation \& Scaling Misuse &
Inappropriate Aggregation; Missing Normalization \\[0em]

Parameter Manipulation &
Arbitrary Thresholds; Post-hoc Parameter Tuning \\[0em]

Statistical Misuse & Insufficient Power; Effect Size Misapplication; Metric Misuse; Uncertainty Omission \\[0em]

\rowcolor{yellow!15}\multicolumn{2}{l}{{\textbf{3. Misleading Visual Encodings}}}\\
Scale \& Geometry Distortion &
Axis/Scale Manipulation; Aspect-Ratio Distortion; Area-as-Quantity; Inconsistent Binning or Tick Intervals; 3-D Perspective Effects \\[0em]

Visual Encoding Misuse &
Color Principle Violations; Improper Chart Types \\[0em]

Perceptual Overload &
Visual Clutter; Over-annotation \\[0em]

Missing/Inaccurate Encoding Annotations &
Missing/Inaccurate Referential Information\\[0em]

Missing Visual Uncertainty &
Omitted Uncertainty Intervals; Concealed Predictive Uncertainty \\[0em]

\rowcolor{green!15}\multicolumn{2}{l}{{\textbf{4. Textual \& Rhetorical Misrepresentations}}}\\
Numeric Inaccuracy &
Incorrect Value Description; Inaccurate Metadata\\[0em]

Rhetorical Overstatement &
Hyperbole; Overstated/Understated Certainty/Severity \\[0em]

Semantic Ambiguity &
Scope Ambiguity; Vague Language; Ambiguous Metric Reference \\[0em]

Visual–Text Mismatch &
Title Slant; Misleading Annotations; Inaccurate Visual Description \\[0em]


\rowcolor{orange!15}\multicolumn{2}{l}{{\textbf{5. Reasoning \& Logical Fallacies}}}\\
Causal Reasoning Errors &
Spurious Causal Inference; Causal Oversimplification; Post-hoc Fallacy \\[0em]

Evidence Distortion &
Misinterpreted/Misused Metric; Misrepresented Source \\[0em]

Faulty Comparison &
Non-Comparable Entities; Denominator Mismatch; Unequal Baseline \\[0em]

Hasty Generalization &
Improper Extrapolation; Overgeneralization/Oversimplification \\[0em]

Predictive Overreach &
Projection as Fact; Unjustified Prediction \\[0em]

Selective Reasoning &
Cherry-picking; Scope Dilution \\[0em]

\rowcolor{cyan!15}\multicolumn{2}{l}{{\textbf{6. Interpretation Biases}}}\\

Perceptual \& Attentional Biases &
Visual Salience Bias; Presentation-Order Effects; Weighted Average Illusion \\[0em]

Inferential \& Reasoning Biases &
Anchoring; Base Rate Neglect; Illusion of Causality; Overgeneralizing Averages \\[0em]

Belief- \& Value-Conditioned Biases &
Confirmation Bias; Belief Bias; Source Credibility Effects \\

\hline
\end{tabular}
\label{tab:taxonomy}
\end{table*}

\subsection{Taxonomy Layers, Definitions, and Cases}

TIC is organized into three layers: high-level dimensions, middle-layer categories, and leaf-level subtypes. The middle layer captures stable mechanisms of breakdown, while the leaf layer remains extensible as new issue patterns emerge.

\begin{tcolorbox}[
    colback=blue!15,
    colframe=blue!60,
    boxrule=0.6pt,
    arc=4pt,
    left=1pt, right=1pt, top=1pt, bottom=1pt
]

\textbf{1. Data-Level Issues.} 
These issues originate before analysis or visualization and compromise the integrity, completeness, or credibility of the underlying dataset.

\end{tcolorbox}

\noindent\colorbox{blue!7.5}{\midlayer{Data Quality Issues}}involve problems in the underlying dataset, such as inaccurate values, missing entries, or misalignment with the intended construct~\cite{mcnutt_2020}. Common cases include \textit{measurement error}, where instruments, transcription, or entry mistakes distort values, and \textit{missing or incomplete data}, where essential records are absent or insufficiently captured, undermining subsequent analysis and visualization.

\par\medskip
\noindent\colorbox{blue!7.5}{\midlayer{Biased/Outdated Data}} issues arise when collected data misrepresent the target phenomenon through biased sampling, structural gaps, or obsolescence~\cite{huff_how_1954,dignazio_klein_2020_datafeminism}. \textit{Sampling bias} occurs when groups are systematically over- or under-represented, as in early COVID-19 case counts that overrepresented hospitalized patients. Structural gaps occur when data collection treats one population as the default, such as datasets built around male bodies and behaviors, producing persistent gender data gaps~\cite{criado_perez_invisible_2019,dignazio_klein_2020_datafeminism}. \textit{Outdated data} refers to data that no longer reflect current conditions but are treated as current.

\par\medskip
\noindent\colorbox{blue!7.5}{\midlayer{Data Absence \& Credibility}}issues occur when claims lack verifiable backing or rely on unreliable sources. \textit{No supporting data} refers to assertions made without citing evidence. \textit{Unreliable data sources} include datasets or publications with unclear origins, opaque methods, or weak domain authority, such as vaccine statistics drawn from anonymous social media accounts or websites that do not disclose collection procedures.

\begin{tcolorbox}[
    colback=red!15,
    colframe=red!60,
    boxrule=0.6pt,
    arc=4pt,
    left=1pt, right=1pt, top=1pt, bottom=1pt
]

\textbf{2. Quantitative Analysis Issues.}
These issues originate when data are transformed into quantitative evidence, including flawed aggregation, arbitrary parameter choices, statistical misuse, and insufficient treatment of uncertainty. \end{tcolorbox}

\noindent\colorbox{red!7.5}{\midlayer{Aggregation \& Scaling Misuse}}arises when grouping, scaling, or baseline choices distort comparisons.

\textit{Inappropriate aggregation}~\cite{huff_how_1954,ge_calvi_2023} combines heterogeneous groups, populations, or time periods in ways that obscure variation or reverse apparent trends. For example, aggregating COVID-19 case counts across distinct waves or policy regimes can be used to claim overall policy success or failure while masking structural breaks. This pattern includes Simpson's paradox, in which relationships observed within subgroups reverse when data are aggregated.

\textit{Missing normalization}~\cite{huff_how_1954,ge_v-framer_2024,lo_misinformed_2022} reports raw counts without adjusting for relevant baselines, such as comparing total COVID-19 cases/ deaths across countries without population normalization, leading to misleading comparisons.

\par\medskip

\noindent\colorbox{red!7.5}{\midlayer{Parameter Manipulation}}skews analytical results through arbitrary or post-hoc parameter choices that reshape how data are constructed, filtered, or summarized, often enabling selective reasoning such as cherry-picking.

\textit{Arbitrary thresholds}~\cite{lisnic_misleading_2023,shilo_visual_2024,alexander_can_2024} introduce cutoffs without empirical or theoretical justification, selectively including or excluding observations in ways that can support a preferred claim. This pattern is especially common in sports commentary, where statistical databases and query tools make it easy to construct claims around eligibility thresholds such as minimum games/minutes played. For example, claims that a player is the ``first in history'' to reach a particular combination of metrics can depend heavily on thresholds selected. 

\textit{Post-hoc parameter tuning}~\cite{wu_computational_2017,walenz_finding_2014} retroactively adjusts analytical settings after inspecting the data, such as time windows, inclusion criteria, or model parameters. This includes practices such as p-hacking and related forms of overfitting, where analytical choices amplify apparent effects rather than reflect stable patterns in the data.

\par\medskip

\noindent\colorbox{red!7.5}{\midlayer{Statistical Misuse}}refers to errors in applying or interpreting statistical methods and outputs that undermine the validity of analytical conclusions~\cite{huff_how_1954,onwuegbuzie_typology_2015,pasquetto_what_2024,mcnutt_2020}.

\textit{Insufficient power} arises when claims are drawn from samples too small to support reliable inference, leading to conclusions that are highly sensitive to noise or natural variability.

\textit{Effect size misapplication} occurs when statistical magnitudes are framed in ways that obscure or exaggerate practical significance, such as emphasizing statistically detectable but substantively small differences without appropriate baseline or contextual information.

\textit{Metric misuse} involves interpreting statistical indicators, such as $p$-values, effect sizes, averages, rates, or model scores, without sufficient context about what they measure and what conclusions they can support.

\textit{Uncertainty omission} occurs when relevant sources of uncertainty, such as variability, measurement error, or model uncertainty, are not estimated, propagated, or otherwise accounted for in the analysis, making quantitative conclusions appear more precise or definitive than warranted. It is closely related to \textit{missing visual uncertainty}, and the two may co-occur when uncertainty that should inform the analysis is also absent from the visual representation.

\begin{tcolorbox}[
    colback=yellow!15,
    colframe=yellow!60,
    boxrule=0.6pt,
    arc=4pt,
    left=1pt, right=1pt, top=1pt, bottom=1pt
]

\textbf{3. Misleading Visual Encodings.} This dimension captures issues in how data are visually mapped, specified, or emphasized, including distortions in scale, geometry, encoding choice, annotation, uncertainty, and visual density. 
\end{tcolorbox}

\begin{figure}[th]
    \centering
    \includegraphics[width=1\linewidth]{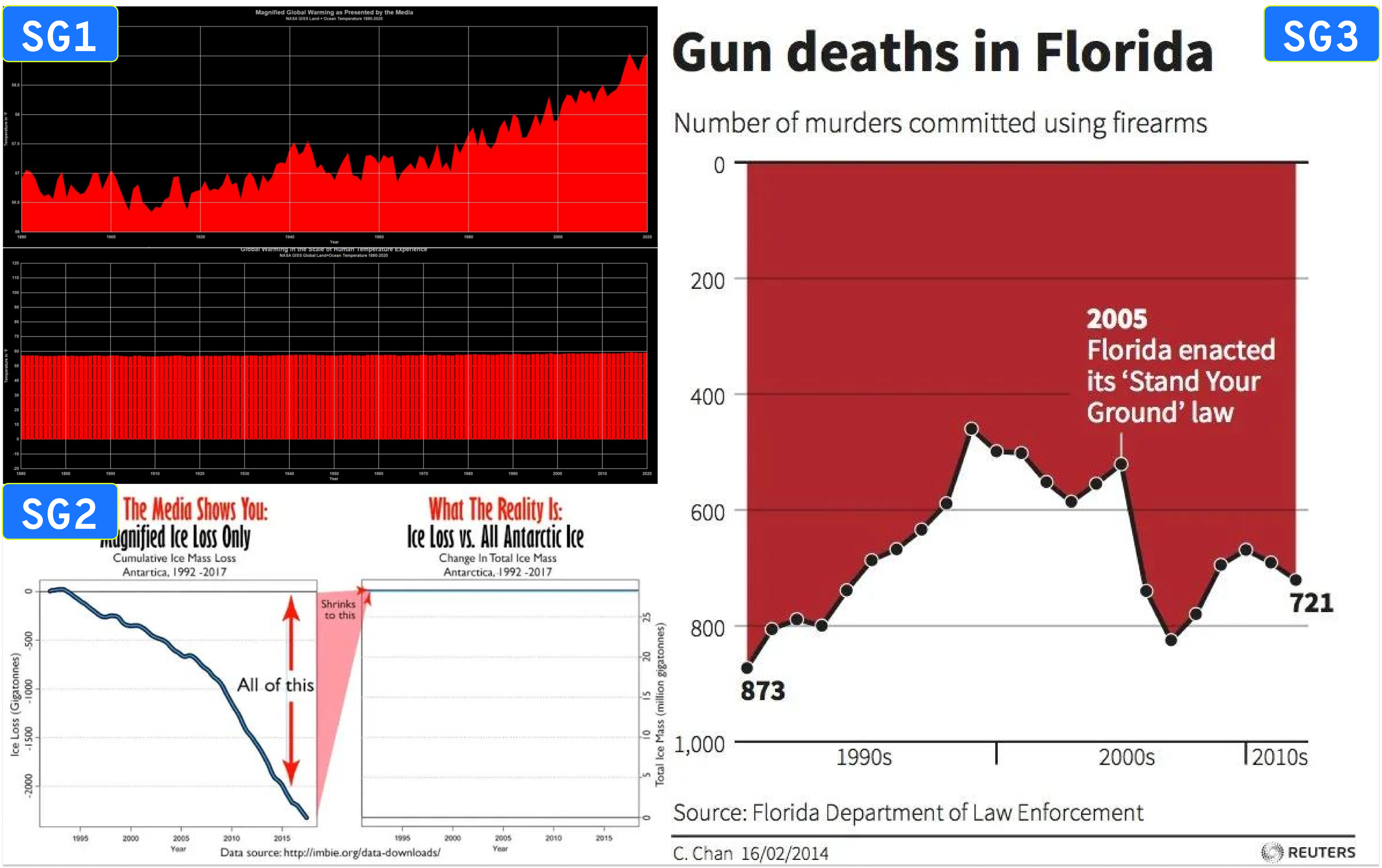}
    \caption{Examples of \textit{Scale \& Geometry Distortions}. 
    \textbf{SG1}: Axis range manipulation compresses or amplifies temperature change by expanding the y-axis to an irrelevant scale. 
    \textbf{SG2}: Scale unit manipulation (millions of gigatonnes) minimizes apparent Antarctic ice loss by shrinking vertical variation. 
    \textbf{SG3}: Axis inversion reverses the visual direction of change, making increases appear as decreases.}
    \label{fig:scale}
\end{figure}

\noindent\colorbox{yellow!7.5}{\midlayer{Scale \& Geometry Distortion}}arises when scaling, axis design, binning, or geometric transformations alter how viewers perceive magnitude, slope, or distribution in ways that can mislead trend or comparison judgments~\cite{howard_wainer_how_1984,pandey_rall_satterthwaite_nov_bertini_2015,lauer_deceptive_2020,lo_misinformed_2022,nguyen_revisiting_2020,ge_calvi_2023,lan_i_2025, correll_truncating_2020}. These techniques are not inherently problematic; they become misleading when they distort the quantitative relationship viewers are expected to assess, or when the transformation is insufficiently justified or explained.

\textit{Axis or scale manipulation} occurs when axis ranges, baselines, directions, or multiple scales exaggerate, compress, or invert apparent differences.

\textit{Aspect-ratio distortion} occurs when chart dimensions are stretched or compressed in ways that alter perceived slopes or trend severity.

\textit{Area-as-quantity} becomes misleading when values are mapped to area or volume in ways that exaggerate, obscure, or make proportional differences difficult to judge.

\textit{Inconsistent binning or tick intervals} occurs when uneven bins, tick spacing, or visually uniform intervals for non-uniform quantities distort perceived distributions or patterns.

\textit{3D perspective effects} become misleading when tilt, depth, occlusion, or perspective cues interfere with accurate judgments of length, area, volume, or relative position.

\par\medskip

\noindent\colorbox{yellow!7.5}{\midlayer{Visual Encoding Misuse}}occurs when visual encodings conflict with data semantics or established perceptual conventions, leading viewers to infer relationships that the data do not support~\cite{howard_wainer_how_1984,nguyen_revisiting_2020,szafir_good_2018,lo_misinformed_2022,lan_i_2025}. These choices are not inherently problematic; they become misleading when the encoding suggests order, proportion, grouping, or contrast that is inconsistent with the underlying data.

\textit{Color principle violations} occur when color choices disrupt the intended data semantics, such as using a non-monotonic or rainbow palette for ordinal or continuous values in ways that introduce spurious boundaries, obscure ordering, or exaggerate contrast. For example, a rainbow color palette for a continuous score may create false thresholds between adjacent values.

\textit{Improper chart types} arise when the chosen visual form is incompatible with the data's relational structure, inviting viewers to make inappropriate perceptual or semantic inferences. For example, using a pie chart for non-partitive relationships, such as comparing average test scores across schools with slices summing to more than 100\%, can invite incorrect part-to-whole reasoning. Similarly, using a line chart to connect unordered categories, such as demographic groups, can imply continuity or temporal progression where none exists.
\par\medskip

\noindent\colorbox{yellow!7.5}{\midlayer{Perceptual Overload}}occurs when visual density exceeds viewers' capacity to efficiently extract relevant patterns, hindering interpretation~\cite{lo_misinformed_2022,alexander_can_2024,lan_i_2025,mcnutt_2020,szafir_good_2018,ge_calvi_2023,ajani_declutter_2022}. Its effects are highly audience-dependent: displays that may be interpretable for domain experts or visually literate readers can overwhelm lay audiences with less expertise, lower data literacy, or limited time and attention. \textit{Overplotting or clutter} from too many marks, and \textit{over-annotation} from excessive labels or callouts, can obscure the main evidentiary pattern. In such cases, visual complexity can increase cognitive load and push viewers toward heuristic processing, making them more likely to defer to the accompanying textual claim rather than independently evaluate the visual evidence.

\par\medskip

\noindent\colorbox{yellow!7.5}{\midlayer{Missing/Inaccurate Encoding Annotations}}occur when textual or graphical specifications needed to decode a visual encoding are absent, incomplete, incorrect, or ambiguous, impairing viewers' ability to interpret how data values are mapped to visual channels~\cite{howard_wainer_how_1984,lo_misinformed_2022,doan_misrepresenting_2021,lan_i_2025}. \textit{Missing or incorrect encoding annotations} arise when legends, axis labels, units, tick labels, or channel descriptions are absent or incorrect, including cases where the stated interpretation of an encoding is wrong (e.g., legends invert the meaning of color, size, or position, or darker colors are described as indicating worse outcomes when they actually represent better values).

\par\medskip

\noindent\colorbox{yellow!7.5}{\midlayer{Missing Visual Uncertainty}}occurs when relevant variability, measurement error, or probabilistic uncertainty is omitted or visually underrepresented, making estimates, forecasts, or comparisons appear more precise than warranted~\cite{lo_misinformed_2022,mcnutt_2020,ge_calvi_2023,shilo_visual_2024,yang_dice_2024}. \textit{Omitted intervals} exclude error bars, confidence bands, margins of error, or ensemble spreads when such uncertainty is necessary for interpreting the claim. \textit{Concealed predictive uncertainty} presents forecasts as point outcomes rather than ranges of plausible scenarios, such as showing only the most likely result in an election forecast. By suppressing uncertainty, such visuals can lead viewers to treat probabilistic evidence as more definitive than it is, increasing confidence in claims that should be interpreted with caution.

\begin{tcolorbox}[
    colback=green!15,
    colframe=green!60,
    boxrule=0.6pt,
    arc=4pt,
    left=1pt, right=1pt, top=1pt, bottom=1pt
]

\textbf{4. Textual \& Rhetorical Misrepresentations.}
This category concerns how data evidence is articulated and framed in text. It includes vague, exaggerated, or misleading language, as well as claims that diverge from the underlying data, metadata, or corresponding visuals.
\end{tcolorbox}

\noindent\colorbox{green!7.5}{\midlayer{Numeric Inaccuracy}}involves quantitative claims that misstate, misattribute, or miscalculate numerical information~\cite{tang_vistext_2023,lan_i_2025,alexander_can_2024,fu_data_2024,vlachos_identification_2015,doan_misrepresenting_2021,cairo_how_2019}, primarily through errors in the reported value itself and, in some cases, through errors in the contextual metadata needed to interpret that value. \textit{Incorrect value description} occurs when the described number is factually wrong, such as when cited/derived values contradict source data, contain computational errors, or reverse the direction or magnitude of a comparison. \textit{Inaccurate metadata description} occurs when a number is attached to incorrect contextual qualifiers, such as the wrong timeframe, geographic scope, population, data source, or sample size, thereby changing what the number represents.

\par\medskip

\noindent\colorbox{green!7.5}{\midlayer{Rhetorical Overstatement}}uses language to make quantitative evidence appear stronger, more certain, more severe, or less consequential than warranted. Although most cases involve exaggeration or intensification, we also include understatement when language minimizes the significance, severity, or uncertainty of the evidence. \textit{Hyperbole} uses emotionally charged or inflated phrasing to heighten perceived importance or danger, such as describing a routine fluctuation as a ``catastrophe'' or a ``disaster.'' \textit{Overstated or understated certainty or severity} occurs when language about magnitude, probability, or impact does not align with the data, for instance, describing a steady increase as a ``surge,'' declaring that an outcome ``will definitely occur'' despite wide uncertainty intervals.

\par\medskip

\noindent\colorbox{green!7.5}{\midlayer{Semantic Ambiguity}}involves imprecise or underspecified language that obscures what is being measured, compared, or concluded. Unlike \textit{numeric inaccuracy}, the issue is not necessarily that a value is factually wrong, but that the claim lacks sufficient specificity for the value or comparison to be interpreted reliably. \textit{Scope ambiguity} arises when key contextual boundaries, such as population, geographic scope, comparison group, or time period, are omitted~\cite{fu_data_2024}, limiting interpretability (e.g., ``Unemployment dropped by 5\%'' without clarifying the group or timeframe). \textit{Vague language} occurs when wording is so indeterminate that it is unclear what data are being referenced or what conclusion is being drawn, such as using terms like ``many,'' ``significant,'' or ``better'' without specifying the basis of comparison. \textit{Ambiguous metric reference} arises when a metric is named or described without essential qualifiers, such as whether a reported change refers to percentage points or percent change, raw counts or rates, or nominal versus inflation-adjusted values. 

\par\medskip

\noindent\colorbox{green!7.5}{\midlayer{Visual-Text Mismatch}}involves misalignments between visual evidence and the text that describes, frames, or annotates it~\cite{Kong2019CHITrustAndRecall,kong_frames_2018,kim_emphasischecker_2024,fu_data_2024,tang_vistext_2023}. \textit{Title slant} occurs when a headline or title overstates, distorts, or selectively emphasizes conclusions not fully supported by the visualization. \textit{Misleading annotations} occur when labels, callouts, or highlights direct attention toward selective or unsupported interpretations~\cite{ge_calvi_2023,lan_i_2025,Fan2022CHI}. \textit{Inaccurate visual descriptions} occur when accompanying text mischaracterizes visual patterns, such as describing a gradual rise as a ``sharp spike'' or claiming divergence when lines actually converge~\cite{kim_emphasischecker_2024,fu_data_2024,tang_vistext_2023}.

\begin{tcolorbox}[
    colback=orange!15,
    colframe=orange!60,
    boxrule=0.6pt,
    arc=4pt,
    left=1pt, right=1pt, top=1pt, bottom=1pt
]

\textbf{5. Reasoning \& Logical Fallacies.}
This category captures flaws in how claims are derived from data, particularly in the reasoning used to construct arguments, interpret relationships, or project outcomes.

\end{tcolorbox}

\noindent\colorbox{orange!7.5}{\midlayer{Causal Reasoning Errors}}arise when causal relationships are asserted or implied without adequate justification, typically by treating correlation, temporal proximity, or partial explanations as causal mechanisms~\cite{huff_how_1954,lan_i_2025,lisnic_misleading_2023,lisnic_yeah_2024,lo_misinformed_2022,alexander_can_2024,onwuegbuzie_typology_2015,kernighan_millions_2018}. These errors are the most prevalent in our corpus and reflect systematic failures to reason about causal structure.

\begin{figure}[th]
    \centering
    \includegraphics[width=1\linewidth]{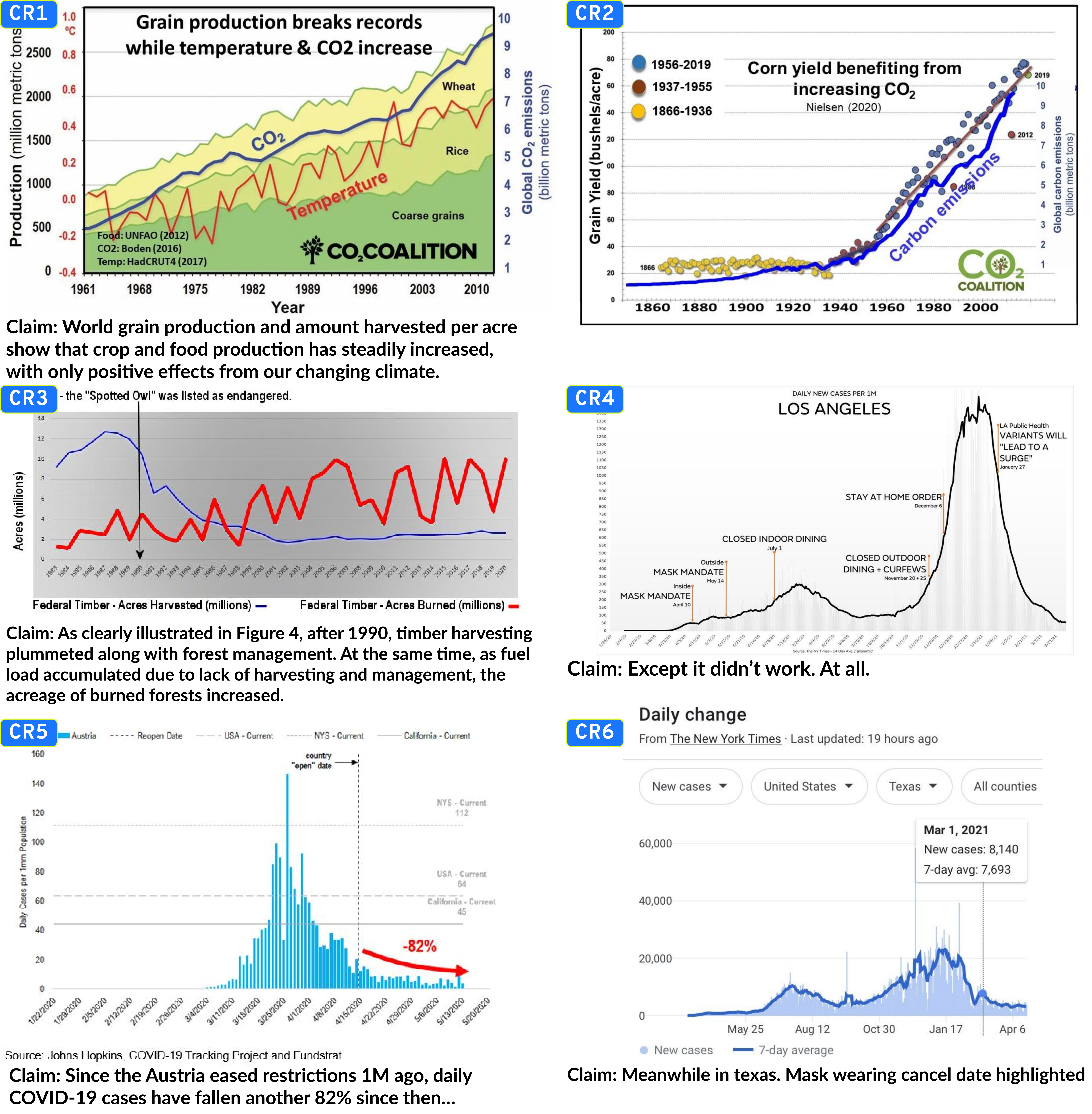}
    \caption{Examples of visual-based \textit{Causal Reasoning Errors}, where correlations or temporal alignments are presented as evidence of direct causation. Cases include claims that rising CO\textsubscript{2} drives crop yields (CR1--CR2), that reduced logging explains wildfire growth (CR3), and that COVID-19 policy shifts directly determine case trends (CR4--CR6).}
    \label{fig:causal}
\end{figure}

\textit{Spurious Causal Inference} treats observed correlation or visual co-variation as sufficient evidence of causation. A common manifestation is visual co-trending, in which two time series are juxtaposed and the accompanying claim implies a direct causal relationship without isolating a mechanism or ruling out alternative explanations. In \autoref{fig:causal}, CR1--CR2 juxtapose agricultural yields with rising CO\textsubscript{2} emissions, while CR3 aligns timber harvesting with wildfire acreage, implying causality primarily from parallel trends.

\textit{Causal Oversimplification} attributes complex outcomes to a single or dominant factor while ignoring confounding variables, mediating mechanisms, and alternative explanations. This pattern frequently co-occurs with spurious inference. For example, the CO\textsubscript{2} narratives in \autoref{fig:causal} omit established drivers of agricultural productivity such as irrigation, fertilizer use, and genetic advances, creating the appearance of a simple, direct causal relationship.

\textit{Post-hoc Fallacy} infers causality from temporal order alone, asserting or implying that one event caused another because it occurred earlier. In \autoref{fig:causal}, CR4--CR6 pair COVID-19 case trends with policy milestones (e.g., reopenings or mandates), framing observed rises or declines as direct policy effects. These narratives often overlook epidemiological delays, confounding factors, and the possibility of reverse causality, where emerging trends may have prompted the policy change rather than resulted from it.

\par\medskip

\noindent\colorbox{orange!7.5}{\midlayer{Evidence Distortion}}refers to claims that use apparently valid evidence while reframing what that evidence can reasonably support. Rather than fabricating numbers, these claims rely on real statistics, metrics, visual evidence, or cited sources but repurpose them to support a different, broader, or stronger conclusion than warranted. The breakdown lies in how the evidence's relevance, scope, or implication is reinterpreted to support a misleading conclusion.

\begin{figure*}[th]
    \centering
    \includegraphics[width=1\linewidth]{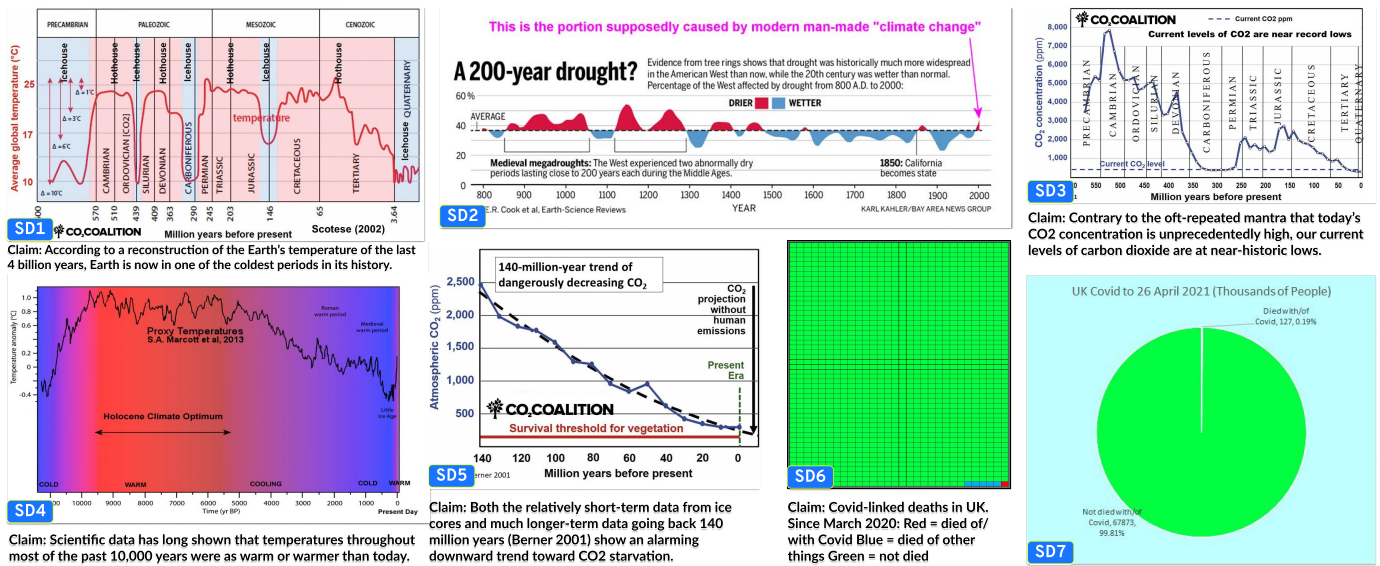}
   \caption{Examples of \textit{Scope Dilution} (a subtype of selective reasoning), where overly broad temporal or spatial baselines---often paired with scale manipulation---are used to downplay present risks. Climate cases (SD1–SD5) invoke geologic or millennial timescales, while public health cases (SD6–SD7) frame COVID-19 impacts against population-level or all-cause mortality baselines.}
    \label{fig:scope_dilution}
\end{figure*}

\textit{Misinterpreted or misused metrics}~\cite{vlachos_identification_2015} occur when technically accurate quantitative indicators are treated as if they measure broader outcomes or risks than they actually do. For example, some treat high voter turnout among registered voters as evidence of electoral irregularities, despite the metric reflecting registration practices and participation rather than fraud. Others use declines in national crime rates to dismiss localized increases. In each case, the metric itself is not incorrect; distortion arises from treating it as a proxy for a different concept than it is designed to measure.

\textit{Misrepresented sources} involve selectively quoting, reframing, or altering cited evidence in ways that change its implications. Common tactics include excerpting findings without their qualifying assumptions, weakening conditional statements, or overlaying annotations onto legitimate charts to redirect interpretation. These strategies preserve the appearance of credibility while substantively distorting the original message.

\par\medskip

\noindent\colorbox{orange!7.5}{\midlayer{Faulty Comparison}}arises when claims rely on invalid contrasts between mismatched groups, baselines, or metrics. Such errors frequently involve overlapping mismatches across populations, denominators, or starting conditions. For example, claims that the U.S. conducted more COVID-19 tests in eight days than South Korea did in eight weeks compare raw testing totals without accounting for population size or testing regimes. Similarly, assertions that SARS-CoV-2 infection rates are higher among vaccinated than unvaccinated populations in the UK rely on incomparable denominators shaped by age structure and baseline risk.
Within this category, \textit{non-comparable entities} compare groups, timeframes, or variables that differ in essential ways~\cite{huff_how_1954}; \textit{denominator mismatch} arises when percentages or rates are computed over differing bases without adjustment~\cite{huff_how_1954}; and \textit{unequal baselines} refer to evaluations of performance or change across groups that begin from markedly different initial levels without normalization~\cite{cairo_how_2019}.

\par\medskip

\noindent\colorbox{orange!7.5}{\midlayer{Hasty Generalization}}arises when claims are applied too broadly or without sufficient evidential support~\cite{huff_how_1954}. One form, \textit{overgeneralization}, treats heterogeneous cases as uniform and ignores important exceptions. A related form, \textit{improper extrapolation}, extends findings beyond the studied context, time range, or population without adequate justification. A third form, \textit{oversimplification}, reduces complex phenomena to reductive explanations that overlook essential variability, mechanisms, or nuance, even when the claim remains within scope. In practice, these errors often co-occur, with claims that both overextend their applicability and collapse complexity into overly simple terms.

\par\medskip

\noindent\colorbox{orange!7.5}{\midlayer{Predictive Overreach}}involves projecting outcomes with unjustified certainty. One form, \textit{projection as fact}, presents hypothetical forecasts or model-based projections as though they were confirmed results. For instance, claiming that ``this policy will reduce crime rates by 50\%'' without providing modeling assumptions, confidence intervals, or sensitivity analyses turns an uncertain forecast into a falsely definitive statement. A related form, \textit{unjustified prediction}, asserts future trends without sufficient data, model validation, or causal mechanisms---for example, predicting that ``vaccine side effects will double over the next five years'' based solely on a short-term fluctuation. These issues often co-occur with \textit{rhetorical overstatement}, where certainty is exaggerated through definitive language.

\par\medskip

\noindent\colorbox{orange!7.5}{\midlayer{Selective Reasoning}}distorts arguments through biased evidence selection, foregrounding supportive data while omitting context that complicates or directly contradicts it.

\textit{Cherry-picking}~\cite{lo_misinformed_2022,lan_i_2025,lisnic_misleading_2023,ge_calvi_2023,lisnic_yeah_2024,pasquetto_what_2024,ge_v-framer_2024} highlights favorable subsets of evidence while suppressing disconfirming data or qualifiers. Across domains including climate communication, public health, education policy, and sports analytics, recurring strategies include \textit{cherry-picking timeframes}, where narrow temporal windows suggest misleading trends; \textit{cherry-picking isolated or extreme cases}, where anomalous reference points or outliers are elevated (often co-occurring with \textit{hasty generalization}); \textit{cherry-picking metrics}, where favorable indicators are emphasized while more comprehensive measures are omitted; \textit{cherry-picking thresholds}, where selectively chosen cutoffs amplify narratives of crisis or success; and \textit{selective demographic framing}, where outcomes for specific groups or locations are highlighted while masking population-level or structural differences.

\textit{Scope dilution}, not explicitly addressed in prior literature, introduces overly broad or weakly relevant context (e.g., expansive temporal, spatial, or population baseline) to downplay, recontextualize, or obscure adverse patterns. As shown in \autoref{fig:scope_dilution}, instances in our corpus, particularly Corpus~III, exhibit this pattern across climate and public health discourse. In climate communication, claims invoke geologic or millennial timescales (e.g., framing the present as ``one of the coldest periods in Earth's history,'' citing near-historic CO\textsubscript{2} levels, or emphasizing centuries-long drought and temperature records), thereby minimizing anthropogenic warming by appealing to fundamentally different planetary conditions. In public health contexts, scope dilution similarly reframes COVID-19 deaths as a small fraction of total population or all-cause mortality. Related strategies compare rising CO\textsubscript{2} to total atmospheric mass or ice loss to total glacier volume. Together, these cases show how scope dilution, often paired with scale manipulation, embeds distortion within misleadingly expansive frames.

\begin{tcolorbox}[
    colback=cyan!15,
    colframe=cyan!60,
    boxrule=0.6pt,
    arc=4pt,
    left=1pt, right=1pt, top=1pt, bottom=1pt
]

\textbf{6. Interpretation Biases.} 
This dimension captures reception-side conditions that shape how audiences interpret data-driven narratives. These biases do not reside in the artifact itself; rather, artifacts may trigger, amplify, or interact with them through visual salience, textual framing, source cues, and evidentiary complexity. They include perceptual attention, cognitive heuristics, data literacy, prior beliefs and values, institutional trust, social identity, and community context, all of which can affect whether narratives are accepted, resisted, or reinterpreted.
\end{tcolorbox} 

Prior work shows that understanding of quantitative information depends not only on data correctness and representational fidelity, but also on perceptual, cognitive, and sociocultural factors (e.g.,~\cite{Kong2019CHITrustAndRecall, dimara_task-based_2020, franconeri_science_2021, lee_viral_2021, Peck2019DataisPersonal, holder_dispersion_2022}). In particular, Dimara et al.~\cite{dimara_task-based_2020} provide a task-based taxonomy of cognitive biases in visualization. Building on prior work, we position these mechanisms within the data communication process, group them into three categories, and highlight how they interact with upstream issues.

\noindent\colorbox{cyan!7.5}{\midlayer{Perceptual \& Attentional Biases}}arise from how visual attention is allocated during initial viewing, shaping which elements are noticed first and how visual evidence is implicitly weighted~\cite{franconeri_science_2021,padilla_decision_2018,baek_ensemble_2020}. These biases reflect constraints of the human visual system in selecting, prioritizing, and summarizing information under limited attentional capacity. One example is the \textit{weighted average illusion}, in which visually salient marks receive disproportionate perceptual weight, biasing the perceived mean of a distribution toward those elements~\cite{hong_weighted_2022,dimara_task-based_2020}. Relatedly, \textit{visual salience bias} and \textit{presentation-order effects} can shape which elements viewers notice first and how strongly they weight them during interpretation~\cite{valdez_priming_2018,franconeri_science_2021}. In practice, these tendencies interact with upstream visual encoding choices: strong cues, such as saturated colors, oversized marks, or prominent annotations, can amplify misleading encodings by directing attention toward features that do not meaningfully reflect the underlying data.

\par\medskip
\noindent\colorbox{cyan!7.5}{\midlayer{Inferential \& Reasoning Biases}}occur when audiences translate perceived data representations into conclusions during active reasoning. These biases encompass cognitive heuristics and mental shortcuts that simplify complex evaluative tasks but can lead to flawed inference~\cite{dimara_task-based_2020,tversky_judgment_1974,xiong_illusion_2019}. Their expression is shaped by data literacy, critical reasoning skills, and domain knowledge, which influence whether viewers interrogate assumptions or accept implications at face value~\cite{ge_calvi_2023}. Representative examples include the \textit{illusion of causality}, in which viewers infer causal relationships from correlational or co-trending patterns and where visual design choices can systematically shift causal judgments~\cite{xiong_illusion_2019}, and \textit{overgeneralization from averages}, where summaries that omit within-group variability encourage viewers to treat a mean as representative of all individuals or subgroups~\cite{holder_dispersion_2022}. Other examples include \textit{anchoring}, where initial values or frames disproportionately shape later judgments, and \textit{base rate neglect}, where viewers overlook relevant background rates or denominators when interpreting a specific claim. These inferential shortcuts interact with issues across dimensions: analytical choices such as missing normalization can promote denominator neglect, while visual encodings that emphasize co-trending patterns can reinforce spurious causal reasoning.

\par\medskip
\noindent\colorbox{cyan!7.5}{\midlayer{Belief- \& Value-Conditioned Biases}}shape how quantitative claims are evaluated, accepted, rejected, or reinterpreted based on prior beliefs, cultural frames, ideological alignment, institutional trust, social identity, and perceived source legitimacy~\cite{lee_viral_2021,Peck2019DataisPersonal}. These biases often emerge when audiences assess credibility, intent, and relevance, particularly in contested domains where data claims are tied to institutions, communities, or lived experience. Representative examples include \textit{confirmation bias}, \textit{belief bias}, and \textit{source credibility effects}, whereby audiences favor claims that align with existing views or defer to trusted messengers, especially when analyses are complex and difficult to independently verify~\cite{lee_viral_2021}. Classic work on \textit{belief bias} shows that people may accept conclusions because they appear intuitively plausible and reject others because they conflict with prior beliefs, even when the underlying logical structure is equivalent~\cite{evans_conflict_1983}. As a result, problematic data claims can remain persuasive despite upstream flaws, because audiences interpret data narratives through relationships of trust, identity, and community alignment, not through evidence alone~\cite{lee_viral_2021,Peck2019DataisPersonal}.

\subsection{Annotated Dataset and Exploration Interface}

\begin{figure}[b]
    \centering
    \includegraphics[width=1\linewidth]{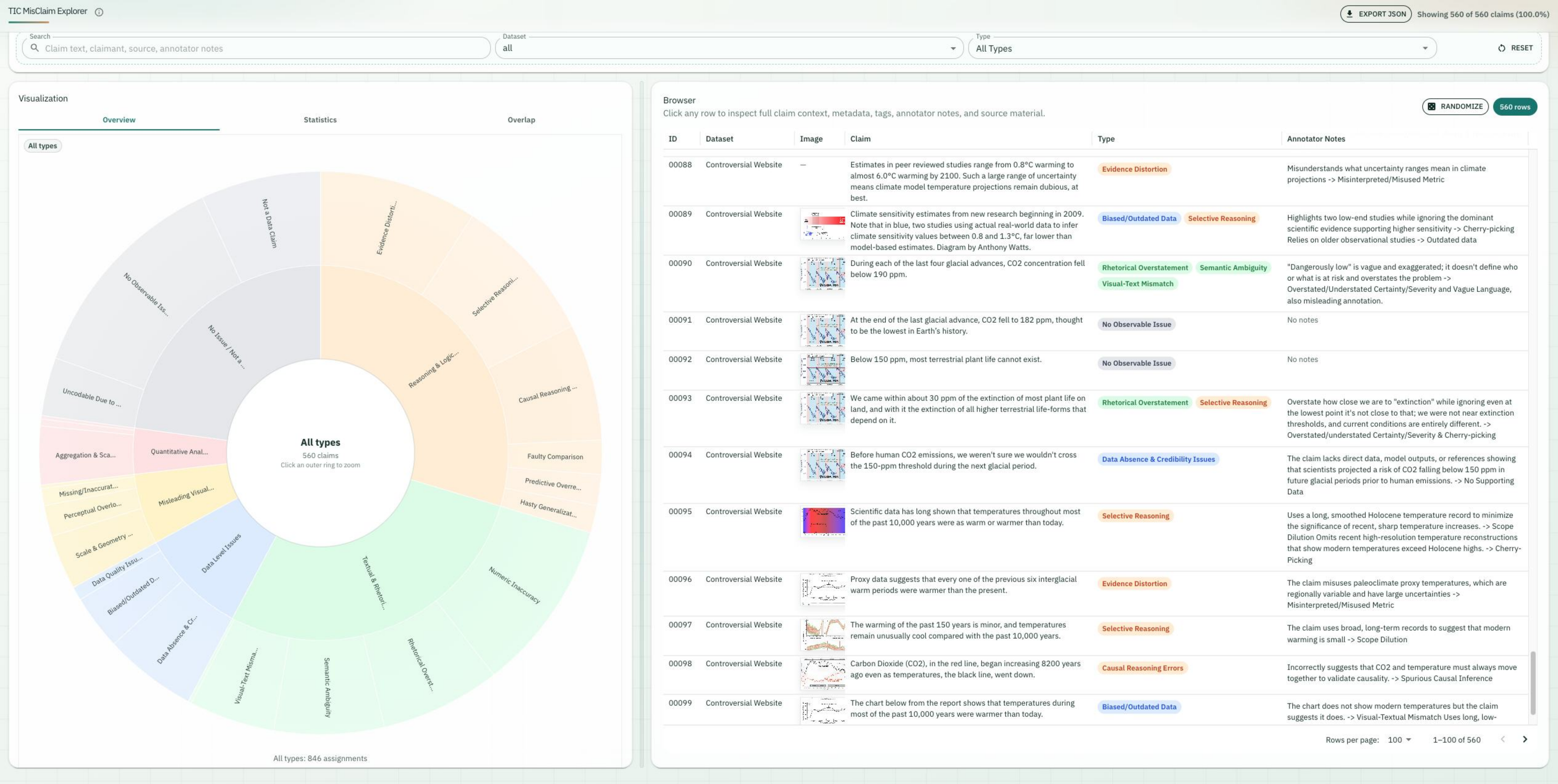}
        \caption{Exploration interface for browsing annotated claims, inspecting issue-type distributions, and reviewing examples with metadata and annotation notes. The interface is available at \url{https://tic-claim.netlify.app/}.}
    \label{fig:interface}
\end{figure}

Applying TIC to the curated corpora yielded an annotated dataset of data-driven claims, issue labels, coding rationales, metadata, and source links. Because TIC supports multi-label annotation, a single narrative may exhibit multiple interacting issue types. To support transparency and closer inspection, we provide an exploration interface (\autoref{fig:interface}) that enables readers to browse annotated claims, examine category distributions and co-occurrence patterns, and review representative examples with their associated metadata, rationales, and sources.

\subsection{Distribution of Annotated Issues}

\begin{figure}[t]
    \centering
    \includegraphics[width=1\linewidth]{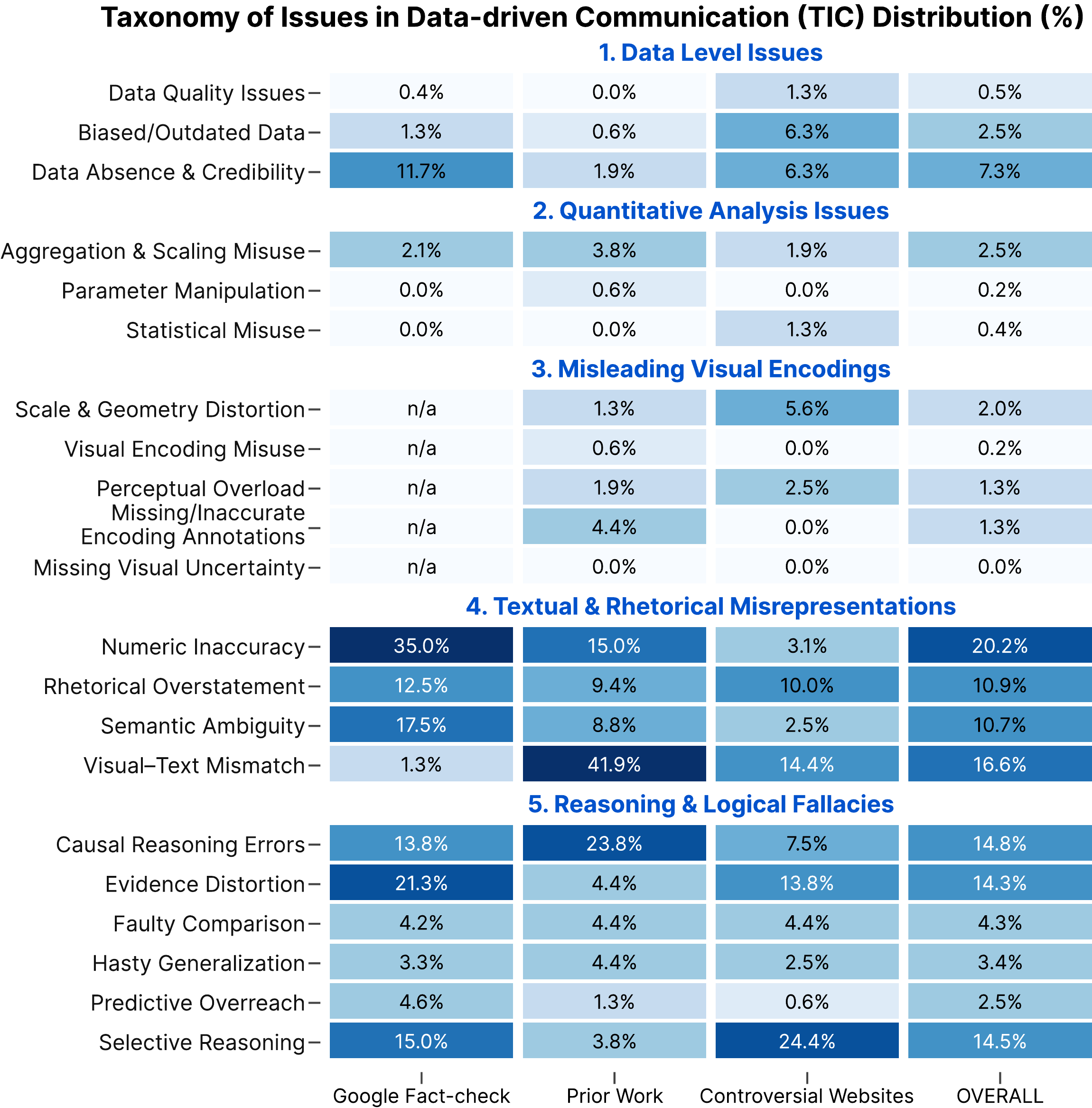}
    \caption{Distribution of annotated TIC issue types across the three corpora. }
    \label{fig:distribution}
\end{figure}

We analyze the distribution of annotated TIC categories to examine how issue types appear across the three corpora. Because TIC supports multi-label annotation, percentages denote the share of narratives within each corpus annotated with each issue type. Interpretation biases are excluded because they are reception-side mechanisms rather than artifact-level annotations. \autoref{fig:distribution} summarizes the resulting distribution.

Overall, \textit{Textual \& Rhetorical Misrepresentations} and \textit{Reasoning \& Logical Fallacies} account for a substantial share of annotated issues, underscoring that data narratives often become misleading through how evidence is articulated, framed, and reasoned about. The most prevalent issue types are \textit{Numeric Inaccuracy} (20.2\%), \textit{Visual--Text Mismatch} (16.6\%), and \textit{Selective Reasoning} (14.5\%), suggesting that problematic data communication frequently arises not only from incorrect numbers, but also from cross-modal misalignment and selective interpretation.

\textbf{Google Fact-Check Corpus} primarily consists of standalone textual claims about quantitative information. In this format, \textit{Numeric Inaccuracy} is especially prevalent (35.0\%), reflecting how incorrect, misattributed, or miscalculated values can be introduced without accompanying data tables or visual context. \textit{Semantic Ambiguity} (17.5\%) and \textit{Rhetorical Overstatement} (12.5\%) further show how vague qualifiers and exaggerated framing can make a claim's scope or evidential strength difficult to assess. Reasoning-related issues, including \textit{Evidence Distortion} (21.3\%), \textit{Causal Reasoning Errors} (13.8\%), and \textit{Selective Reasoning} (15.0\%), indicate that many textual claims extend beyond reporting quantities to advance broader interpretive or causal arguments.

\textbf{Prior Work Corpus} focuses on claims made about visualizations. Although misleading visual encodings occur, the dominant issue is \textit{Visual--Text Mismatch} (41.9\%), where captions, headlines, or narrative descriptions extend beyond what the visualization supports. In many cases, the visual provides descriptive or correlational evidence, while the surrounding text introduces stronger claims, such as generalizations or causal explanations. This pattern suggests that misleading visualization-based narratives often arise not from chart design alone, but from how visual evidence is described, framed, and reasoned about. The high prevalence of \textit{Causal Reasoning Errors} (23.8\%) further shows how visual patterns are frequently used to justify causal conclusions even when the data support only association, comparison, or trend description.

\textbf{Controversial Websites} frequently engage with high-stakes, value-laden domains such as climate and public health. In this corpus, misleading narratives less often depend on overt numerical inaccuracies and more often rely on selective or reframed uses of otherwise plausible evidence. \textit{Selective Reasoning} is especially prevalent (24.4\%), including strategies such as cherry-picking and scope dilution that shape interpretation while preserving the appearance of factual legitimacy. In many cases, valid data, statistics, or sources are cited but selectively chosen, incompletely contextualized, or framed against inappropriate baselines. The prevalence of \textit{Evidence Distortion} (13.8\%) and \textit{Visual--Text Mismatch} (14.4\%) further indicates that misleading narratives in this corpus are often sustained by reinterpretation and reframing of evidence rather than by easily falsifiable numerical errors.

\section{Situating TIC in Data Communication Process}
\label{sec:framework}

TIC identifies recurring issues in data-driven communication, but its broader value lies in showing where those issues enter and propagate through the communication process. To support this process-level view, we situate TIC across the stages through which data are transformed into narratives, visual representations, and public interpretations. Prior frameworks have provided valuable insights into the data storytelling process~\cite{lee_more_2015, li_where_2024} and visualization process models~\cite{card_readings_1999}, but they tend to emphasize the construction of data artifacts, giving comparatively less attention to the upstream and downstream stages of communication, as well as to the finer-grained meaning-making activities that unfold across the two primary modalities of text and visualization.

Drawing on these frameworks and Hall's encoding/decoding model~\cite{hall_encoding_1980}, which frames communication as a process in which producers encode meaning into artifacts and audiences decode it through their own interpretive frames, we conceptualize data communication as a process of meaning construction rather than a simple transfer of information. In this process, meaning is shaped through analytic choices, narrative and visual construction, audience interpretation, and recirculation. Guided by this lens, we organize the data communication process into three stages: \textit{analysis}, where evidentiary possibilities are shaped through data collection, processing, analysis, and insight selection; \textit{construction}, where findings are encoded into multimodal artifacts; and \textit{consumption}, where audiences interpret, evaluate, and recirculate these artifacts. Encoding begins in analysis and becomes explicit in construction, while decoding takes place in consumption, where narratives may be accepted, negotiated, or contested. We introduce each stage in turn, outlining its activities, associated issues, leverage points for trustworthy practice, and actors involved.

\subsection{\textit{Analysis}: Extracting Meaning From Data}
The analysis stage establishes the foundation of the process, where raw phenomena are captured, structured, and analyzed. Decisions about what is measured, how it is processed, and which methods are applied inscribe assumptions into the dataset, shaping all downstream narratives.

\textbf{Activities.} 
This stage begins with \textit{data collection}, which may take multiple forms: raw traces generated by \textit{instrument designers} (e.g., sensors, logging systems, automated pipelines), datasets assembled by \textit{data providers} (e.g., platforms or companies), or data directly gathered by \textit{authors and analysts} through surveys, scraping, or fieldwork. Next, \textit{data processing and cleaning}, typically handled by authors, resolve missing values and inconsistencies while structuring data for analysis. Finally, \textit{exploration and modeling} use methods such as descriptive statistics, visual analysis, and predictive models to surface candidate patterns.

\textbf{Issues.} Decisions at this stage are where the first two TIC dimensions often emerge. During collection, \textit{data integrity issues} may arise from measurement errors, transcription mistakes, or biased samples that exclude certain populations. Collection is also a common source of \textit{representativeness and provenance issues}, for example, when outdated datasets are used, sources lack transparency, or citations are missing. Processing can further degrade provenance when contextual metadata is discarded or transformations are undocumented. Modeling frequently introduces \textit{analytical misuses}, including aggregation that conceals variation, ill-defined comparison groups, or reliance on unnormalized counts. Once embedded, these issues can cascade downstream into construction and consumption, where they become harder to detect or repair.

\textbf{Leverage points.}
At this stage, leverage lies in helping actors make the evidentiary basis of a narrative auditable before it enters construction. \textit{Data providers} can be supported in clarifying provenance, measurement assumptions, collection scope, metadata, missingness, uncertainty, and metric definitions, including how key metrics are computed and should be interpreted. \textit{Authors/analysts} can be supported in making analytical transformations inspectable, including cleaning, filtering, aggregation, normalization, parameter choices, and modeling assumptions. These traces help \textit{editors or fact-checkers} assess whether the data adequately represent the target phenomenon, whether important contexts are missing, and whether the analysis can support the intended claim.

\subsection{\textit{Construction}: Embedding Meaning in Artifacts}
The construction stage is where analytic findings are transformed into the communicative artifacts--- \textit{data-driven narratives}. Authors move from a pool of potential insights to a framed narrative by selecting evidence, articulating claims, developing reasoning, and designing visualizations.

\textbf{Activities.} Construction unfolds through a series of compositional moves. Authors first select which insights to foreground from a broader pool that may include both supporting and contradictory evidence. These insights are articulated as \textit{data claims}, ranging from factual descriptions to scoped comparisons across values, groups, or time. Claims may be extended through narration and interpretation, where authors explain patterns, assign meaning, or project implications, and further developed into arguments that link evidence to broader conclusions. In parallel, authors construct visual elements by designing charts, adding annotations, and framing visualizations to emphasize selected takeaways. Narrative visualizations~\cite{segel_heer_2010} intentionally integrate plots, highlights, and text to guide interpretation and focus audience attention.

\textbf{Issues.} This stage is where textual, visual, and reasoning issues in TIC most often arise. During claim articulation, textual and rhetorical problems can include numeric inaccuracies, semantic ambiguity, and overstatement of certainty or severity. Visual representations may distort meaning through scale or geometry choices, encoding misuse, perceptual overload, missing uncertainty, or misleading annotations~\cite{lo_misinformed_2022}. At the reasoning level, communicators may overreach causally, misinterpret evidence, compare non-comparable groups, or selectively foreground favorable data. Together, these issues show that construction is not a neutral translation of evidence into narrative, but an active process in which meaning can be selected, amplified, reframed, or distorted.

\textbf{Leverage points.} 
During construction, leverage lies in helping authors, editors, and reviewers preserve the connection between selected evidence, visual representation, textual framing, and reasoning. Authors can be supported in reviewing evidence selection, including whether chosen values, time windows, baselines, comparisons, examples, or omitted counterevidence create a selective account. They can also be supported in checking semantic alignment across data, visuals, and text, and in making the reasoning behind claims more inspectable, especially when claims move from description to interpretation, causality, or prediction. Editors and internal reviewers can use these traces to assess whether the narrative maintains fidelity to the underlying evidence, whether visual and textual meanings remain aligned, and whether the argument from evidence to conclusion is coherent while revision is still possible.

\subsection{\textit{Consumption}: Decoding and Circulating Narratives}
The consumption stage concerns how data-driven narratives are interpreted, appropriated, and reshaped by audiences. Once artifacts circulate, their meaning is no longer fixed; it is actively decoded through audiences’ prior knowledge, beliefs, values, and cognitive heuristics.

\textbf{Activities.} Audiences encounter and interpret narratives by drawing on data literacy, prior knowledge, lived experience, institutional trust, community norms, and trusted sources to assess significance and credibility. As they engage with text and visuals, framing and ordering guide attention, shaping which elements appear most salient. Interpretation is further filtered through belief systems and social identities, leading audiences to accept, negotiate, or reject claims~\cite{hall_encoding_1980}. In digital environments, consumption often extends to reproduction: narratives may be reposted, remixed, or recontextualized with commentary, circulating in forms increasingly detached from their original framing.

\textbf{Issues.} This stage foregrounds the final dimension of TIC, \textit{audience interpretation biases}. Heuristic shortcuts can distort judgments of importance or reliability, while attentional biases are shaped by visual emphasis, framing, and presentation order. Belief-driven interpretation further filters claims through worldviews and trust networks, producing selective uptake or rejection~\cite{hall_encoding_1980}. Digital circulation can amplify these effects, as narratives are reframed or stripped of context to support alternative agendas. When audiences actively reconstruct or reinterpret narratives, they effectively assume an authoring role, potentially reintroducing issues from earlier stages, including textual misrepresentation, flawed reasoning, or misleading visual encodings.

\textbf{Leverage points.} 
Once narratives circulate, leverage shifts from preventing breakdowns to helping actors recover context, inspect evidence, and evaluate how interpretation is being shaped. \textit{Fact-checkers and verification organizations} can be supported in reconstructing a claim's evidentiary context and plausible alternative interpretations, and in translating this reconstruction into evidence-linked explanations of why a narrative may be inaccurate, incomplete, or misleading. \textit{Platforms and interface designers} can support this process by preserving source context, surfacing provenance and uncertainty cues, and connecting claims to relevant evidence without presenting automated judgments as final verdicts. Because audiences are socially situated interpreters, downstream support should help them compare interpretations and contextualize claims through trusted sources and communities, rather than treating misunderstanding as only an individual literacy problem.

\par\medskip

By aligning TIC categories with activities, actors, artifacts, and leverage points, the framework clarifies where issues arise and what can be leveraged to address them: auditable evidence and analysis, evidence-to-narrative fidelity in construction, and restored evidentiary context in consumption. These leverage points motivate the validation strategies and sociotechnical supports discussed in the next section.

\section{Discussion}

Building on TIC and our data communication framework, we discuss four questions for understanding and addressing problematic data communication: how data narratives should be assessed as communicative packages, how authorial intent should be treated, how scrutiny should be calibrated across contexts, and how technological support should be designed.

\subsection{Assessing Data Narratives as a ``\textit{Whole Package}''}

A critical question in evaluating data-driven narratives is: \textit{What should be validated, and against what?} TIC surfaces three forms of validation across the communication process: \textit{evidence-grounded assessment}, which checks the narrative against external data, sources, provenance, and uncertainty; \textit{internal alignment assessment}, which examines whether data, visuals, text, annotations, and claims cohere within the artifact; and \textit{contextual/inferential assessment}, which asks whether the framing, comparison structure, causal language, and omitted context support the interpretation being invited. These forms of validation correspond to checkpoints across TIC dimensions: whether the underlying data are valid, representative, current, and credible; whether transformations, metrics, baselines, and evidence selection are appropriate; whether visual encodings faithfully represent the data and fit the semantic meaning of the insight; whether text accurately describes the data and visuals; and whether comparisons, causal claims, predictions, and generalizations are warranted. Because these checks span the full communicative process, TIC suggests that evaluation should target the whole communicative package rather than isolated numbers, visual encodings, or factual claims.

These checkpoints should be interpreted as an interacting chain of checks rather than as independent verdicts. A data narrative's meaning emerges not only from whether each component is accurate in isolation, but from how components reinforce, qualify, or counteract one another across the communicative package. TIC therefore supports diagnosis rather than binary classification: it helps identify where communicative breakdowns occur, how they compound, and what forms of repair---additional context, corrected wording, revised encoding, or reframed inference---may be needed. The Florida gun-death visualization in \autoref{fig:scale}, for example, uses an inverted vertical axis: larger numbers appear lower on the chart, contrary to the conventional ``higher means more'' reading. This choice confused many viewers, who interpreted the post-law period as showing a decline in gun deaths. Yet the designer reportedly described the choice as showing deaths in negative terms, inspired by a dripping-blood visual metaphor~\cite{mills2024truth}. Paired with a textual description such as ``\textit{The chart uses a downward, blood-like metaphor to show that gun deaths increased after Florida enacted its Stand Your Ground law},'' the visual form remains risky but is semantically aligned with the data. Paired instead with ``\textit{Gun deaths declined after the law was introduced},'' the narrative then becomes misleading because the text reverses the direction of change. This example illustrates why validation should synthesize component-level checks into an assessment of the whole communicative package, rather than treating any single component as independently decisive.

\subsection{The Role of Authorial Intent in Assessing Data Narratives }

A complementary question concerns the role of authorial intent: \textit{How should intent inform the assessment of a narrative?} We argue that intent is meaningful but difficult to operationalize. It is often invoked to distinguish benign error from deliberate deception, yet this binary can obscure the complexity of real-world communicative practice. Authors may seek to inform, advocate, persuade, educate, shape interpretation, or maximize engagement, often while navigating editorial constraints, platform incentives, institutional goals, and iterative design feedback. Persuasion and advocacy are not inherently problematic; they become communicatively problematic when rhetorical framing, evidentiary selection, or inferential structure invites interpretations that exceed, obscure, or reverse what the data can reasonably support. Because intent is rarely recoverable as a single, stable motive from the artifact alone, treating it as the primary basis for evaluation risks shifting attention away from the observable properties of the communication itself.

Intent is also not directly observable in an artifact. Instead, it is inferred from rhetorical cues, signals of stance, credibility, and audience appeal. In this sense, intent is interpreted through dimensions commonly associated with \textit{ethos} and \textit{pathos}~\cite{braet_ethos_1992, higgins_ethos_2012}. These inferences are inherently relational: they depend not only on the author's rhetorical choices but also on the audience's prior beliefs/stances and interpretive framework. A reader skeptical of a source may engage in motivated skepticism~\cite{taber_motivated_2006}, interpreting a minor design choice as deliberate misrepresentation, whereas a more trusting audience may attribute the same feature to benign error or technical constraint. Because judgments about intent emerge through this interaction between authorial projection and audience reception, treating intent as a primary evaluative criterion risks amplifying subjectivity rather than resolving it, and may limit agreement among evaluators with differing interpretive frameworks.

TIC addresses these challenges by grounding assessment primarily in observable properties of data narratives. Most TIC dimensions foreground \textit{logos}: the evidence, analytical choices, representational decisions, textual claims, and reasoning structures that can be directly inspected and compared~\cite{braet_ethos_1992, higgins_ethos_2012}. At the same time, TIC is artifact-centered but not artifact-only. Its interpretation-bias dimension captures the reception side of data communication, where audience attention, prior beliefs, values, and source perceptions shape how narratives are understood. These biases may be triggered or amplified by artifact features, but they are not fully contained in the artifact itself. The goal, therefore, is not to determine whether an author meant to deceive, but to assess whether a narrative is erroneous, malformed, misrepresentative, or likely to support misleading interpretation. This approach provides a more stable and operationalizable foundation for assessment while treating intent and audience interpretation as complementary contextual dimensions rather than primary evaluative criteria. At the same time, TIC does not preclude inquiry into intent. By making problematic narrative properties visible and analyzable, it establishes an empirical basis for studying how authorial goals, production constraints, audience predispositions, and dissemination contexts shape the emergence and reception of problematic data narratives.

\subsection{Contextual Calibration of Scrutiny}

Another question is \textit{how scrutiny should be calibrated}. TIC issues are not equally consequential across contexts: their salience, likely harms, and appropriate responses depend on stakes, authority, audience vulnerability, and communication format. Calibration does not mean lowering evidentiary standards or changing what counts as a problem. Rather, it means prioritizing which issues require closer scrutiny, richer contextual explanation, faster correction, or more explicit uncertainty communication in a given communicative setting.

One factor is \textit{stakes and authority}. In high-stakes settings, such as scientific reporting, public policy, and public health, even subtle issues---such as \textit{omitted uncertainty}, \textit{unclear metric definitions}, or \textit{overstated certainty}---may produce substantial downstream harm and therefore warrant closer scrutiny. Scrutiny is also shaped by who authors, conveys, or amplifies a narrative. Claims from public officials, political leaders, major media outlets, or scientific institutions require careful evaluation because their reach and perceived legitimacy can magnify potential impact. Conversely, informal or low-authority claims may still warrant close scrutiny when they circulate widely or are amplified by high-authority actors.

A second factor is \textit{audience vulnerability}. Calibration should account for how readily the intended or likely audience can recognize and evaluate a potential issue. The same \textit{misinterpreted metric}, \textit{faulty comparison}, or \textit{denominator mismatch} may warrant stricter scrutiny when audiences lack the domain knowledge, numeracy, or access to underlying data needed to assess it. In such cases, evaluators should place less tolerance on implicit assumptions, undefined metrics, missing baselines, or unexplained uncertainty, because the narrative itself carries more responsibility to make these conditions explicit.

A third factor is \textit{communication format}. Spoken, written, and visual narratives afford different levels of qualification, sourcing, and correction. Spoken narratives are often produced under temporal and cognitive constraints, increasing the likelihood of issues such as \textit{inaccuracy}, \textit{ambiguity}, and \textit{overstatement}. Written and visual narratives, by contrast, more readily support sourcing, contextual elaboration, and revision, and therefore invite stronger expectations for narrative integrity. These differences do not exempt any format from scrutiny; rather, they inform how evaluative standards should be prioritized.

\subsection{Design Implications for Technological Support}

A final question concerns \textit{how technological support should be designed to address TIC failures}. The leverage points in the process indicate where different actors are best positioned to prevent, detect, or mitigate such failures, and thus where support can be most effective. Because different stakeholders operate under distinct constraints, interventions should be role-specific rather than uniform. Here, we outline design implications aligned with these leverage points.

\subsubsection{Upstream Authoring and Editorial Support}
At the \textit{encoding} stage, analytical findings are transformed into data-driven narratives. Technical interventions here can address upstream issues before publication by helping authors and editors produce clearer, more accurate representations while reducing tedious or error-prone work.

\textbf{Visualization Design Feedback and Linter.}
A visualization linter can be embedded in authoring workflows to provide real-time checks over chart specifications or rendered output, often combining rule-based heuristics with data-aware validation and, in some cases, automatic fixes or in situ annotations~\cite{hopkins_visualint_2020,chen_vizlinter_2022,lei_geolinter_2024, mcnutt_linting_2018}. Complementary systems provide post hoc critiques of completed charts, for instance via perceptual simulation or LLM-based review, supporting iterative refinement and assessments of clutter or unnecessary complexity~\cite{shin_visualizationary_2025}. A continuing challenge is producing actionable guidance for highly customized or domain-specific visualizations, where appropriate complexity depends on audience and communicative goals.

\textbf{Data Narration Assistance.} 
Tools in this category support authors in transcribing evidence from spreadsheets, analyses, or visualizations into narrative text. Prior work spans NLG approaches such as \textit{data-to-text generation}~\cite{wiseman-etal-2017-challenges, parikh-etal-2020-totto} from structured sources (e.g., tables and knowledge graphs)~\cite{lin_survey_2024} and \textit{visual-based NLG} that generates text from visualizations for chart understanding and automated storytelling~\cite{hoque_natural_2025}. While these approaches can accelerate authoring, they risk factual drift, bias, or misleading phrasing and may miss the nuance or framing intended by human authors. More promising are \textit{interactive authoring tools} that support authors before, during, and after writing: surfacing candidate claims and salient patterns, offering data-grounded autocomplete and phrasing suggestions~\cite{Chen_2022_CrossData,fu_dataweaver_2025}, and prompting clarification, precision, and missing context during revision.

\textbf{Claim Review \& Revision.}
Support can also occur during \textit{review and revision}. Tools can help authors/editors refine claims along several dimensions. First, systems can verify \textit{numeric accuracy} by checking whether reported numbers match underlying sources or recomputed results~\cite{fu_data_2024}. Second, systems can check \textit{visual-text alignment} to ensure that textual descriptions match the corresponding visualization in trends, magnitudes, and comparisons (e.g.,~\cite{kim_emphasischecker_2024}). For instance, if a line chart shows cases increasing while the text describes a ``sharp decline,'' the system can flag the mismatch. Third, systems can assess \textit{context completeness} by prompting authors to specify critical framing details (e.g., timeframe, geographic scope, population, units). Finally, systems can improve \textit{semantic clarity} by flagging vague, ambiguous, or overstated language and recommending more precise phrasing grounded in the data and interactions. Together, these tools can function as collaborative checks that preserve authorial control while reducing cognitive load and improving accuracy and contextualization.

\textbf{Reasoning and Argumentation Support.}
Authors may introduce reasoning flaws that mislead readers even when the underlying data are correct. Tools providing \textit{data reasoning guardrails} can flag possible signs of \textit{cherry-picking}, unsupported \textit{causal links}, and overlooked \textit{confounding factors}, while prompting authors to reconsider causal statements inferred from correlational data. Beyond flagging such issues, systems can offer \textit{reasoning guidance} that helps authors examine whether claims are adequately supported, what assumptions they rely on, and what alternative explanations remain plausible, e.g., encouraging distinctions between correlation and causation, clarification of comparison baselines, or acknowledgment of uncertainty and scope conditions. Complementing this, \textit{gentle nudges} can introduce lightweight friction at key moments in the writing process (e.g., prompting reflection when strong causal language is used or when relevant variables may be omitted), supporting more careful reasoning without constraining author agency. Finally, \textit{argumentation scaffolding}, drawing on frameworks such as Toulmin's Model of Argumentation~\cite{toulmin2003uses}, can guide authors in structuring claims, making warrants explicit, qualifying conclusions, and considering counterarguments, thus making the inferential structure of data narratives more transparent during construction.

\subsubsection{Downstream Evaluation Support}

Many downstream issues mirror those arising upstream; the key difference is often less what can be detected than what evidence is available to adjudicate it. Fact-checkers and audiences in particular face a central constraint: limited access to original, reliable, and standardized reference data.

\textbf{Data Infrastructure.}
Downstream verification depends on accessible public data and standardized APIs that enable efficient querying, cross-referencing, and validation by both automated systems and human fact-checkers. Public initiatives (e.g., \textit{Open Government Data} platforms, the U.S. Census Bureau, and the Bureau of Labor Statistics) provide structured datasets across domains such as demographics, labor markets, and public health. Some ecosystems---notably professional sports (e.g., NBA, MLB)---offer near-real-time, high-granularity statistics via public-facing APIs, supporting transparency, third-party analysis, and audience engagement. Such infrastructure enables verification pipelines to link narrative claims to trusted references, but many domains still face limited availability, fragmented sources, and weak standardization. For private or hard-to-access data, third-party platforms may help standardize retrieval and support consistent, secure access for verification.

\textbf{Fact-Checking Reporting Aid.}
Effective downstream support should not only assess claim accuracy but also streamline the production of evidence-linked justifications. Fact-checkers must often combine visual context (e.g., trend comparisons, full distributions), critical framing details (e.g., timeframe, population), and external evidence (e.g., public datasets, scientific studies). Reporting tools can scaffold this work through modular templates that populate claim components, retrieve relevant sources, and generate appropriate visual slices. For example, when a claim about unemployment omits months, the system could produce a time series covering the full period. Interactive report builders can further support annotation, citation linking, and structured write-ups that explain why a claim is inaccurate, incomplete, or misleading, reducing both time and cognitive effort.

\textbf{Issue Alert Representation and Communication.} 
For audiences, support should make potential issues visible and interpretable without reducing evaluation to a simple verdict. Simple techniques such as color-coded text can immediately signal the nature of a problem~\cite{fu_data_2024}. More advanced alert systems can generate automated charts to visualize missing or misrepresented data. For example, if a claim is flagged for cherry-picking~\cite{lisnic_misleading_2023,Lisnic_2025_VisualGuardrail}, the system could display the excluded data points alongside the selected subset to provide context. For causal overreach, it could surface evidence of confounding factors (e.g., alternative trends) or show confidence intervals to highlight uncertainty. Alerts may also be layered with severity indicators (e.g., critical, moderate, minor) and presented through overlays, sidebars, or interactive highlights that let readers explore supporting evidence. Designing these alerts to balance visibility, clarity, and cognitive load is essential to empower both fact-checkers and the public to scrutinize and engage with data narratives.

\subsubsection{Potential and Risks of Using LLMs for Evaluation}
LLMs can support evaluation by surfacing background knowledge that is not readily available from the immediate data or narrative. They can clarify specialized metrics, provide domain context, and make implicit assumptions explicit. For example, when assessing claims such as ``Earth was much warmer 300 million years ago, so modern climate change is irrelevant,'' an LLM can supply paleoclimate context and emphasize that contemporary concern centers on the pace of change and anthropogenic drivers. In sports, LLMs can explain limitations of common metrics (e.g., PER) and highlight omitted dimensions such as defense or role effects. They can also flag plausible confounders that undermine causal readings (e.g., the lighter-lung cancer association without accounting for smoking). With multi-modal capabilities, LLMs may further help detect visual-text inconsistencies and common visual pitfalls (e.g., truncated axes, cherry-picked comparisons)~\cite{alexander_can_2024}.

However, LLM-based evaluation carries nontrivial risks. Models may hallucinate statistics, misstate quantitative relationships, or produce plausible-sounding but incorrect explanations, especially when multi-step reasoning or statistical nuance is required. Because LLM outputs often lack transparent links to primary data and explicit computational steps, errors can be difficult to verify and easy to miss. Accordingly, LLMs should be treated as assistive tools that support, rather than replace, human verification.

\section{Limitations}
The preceding sections present TIC as a process-oriented account of issues in data communication. Here, we clarify the main constraints that shape how this account should be read and where future work can extend it.

\subsection{Dataset Biases in Domain and Modality}
Our dataset centers on high-profile domains such as climate change, COVID-19, economics, and politics, reflecting the priorities and coverage patterns of fact-checking platforms, prior research datasets, and controversial websites. While this captures many recurring topics and data types in public data communication, it does not provide exhaustive coverage; our goal is to identify recurring issue patterns rather than estimate their prevalence across all data narratives. Coverage is also constrained by modality: we analyze primarily static formats while excluding dynamic forms such as dashboards, animations, audio, video, and scrollytelling. Future work should examine how interactivity and dynamic affordances introduce distinct biases and communicative risks, and should extend to additional domains to uncover more fine-grained, domain-specific issues, such as the unique challenges of communicating healthcare and medical data.

\subsection{Analytical Choices and Subjectivity}
We do not report formal inter-coder reliability for the final corpus because our goal was formative taxonomy development rather than producing mutually exclusive ground-truth labels. Many narratives involved overlapping issues across dimensions, where coding differences often reflected boundary judgments or differences in granularity rather than simple error. To support analytic dependability, we used pilot coding, independent subset annotation, reconciliation discussions, iterative codebook refinement, documented rationales, and spot-checking by companion coders. Future work should further assess TIC's reliability and usability through larger-scale independent coding or expert validation.

\subsection{Category Boundaries and Issue Propagation}
TIC categories should be interpreted as analytic lenses for locating and characterizing breakdowns, rather than as mutually exclusive error classes. Because data narratives unfold across a communication process, a single problem may appear at multiple points: choices in data collection, analysis, or evidence selection can later shape visual encoding, textual framing, and reasoning. For example, missing normalization may underlie a faulty comparison, selective evidence may be reinforced through visual emphasis, and weak causal evidence may be expressed as rhetorical overstatement or causal overreach. Thus, apparent overlap among categories often reflects how issues propagate and manifest across stages, not redundancy in the taxonomy. While TIC helps diagnose where and how a narrative breaks down, it does not by itself establish a complete causal account of how one issue produced another.

\subsection{Scope of the Decoding Phase and Reproduction}
While TIC draws on Hall's encoding/decoding model to conceptualize data communication, it does not fully examine downstream feedback, reinterpretation, and reproduction. As audiences repurpose data narratives, issue prevalence may shift and new breakdowns may emerge. While TIC provides useful starting points for analyzing these processes, operationalizing the consumption phase requires shifting the focus from artifacts to audience practices, platform dynamics, and circulation contexts. We therefore call for future empirical work (e.g., \cite{lisnic_yeah_2024}) to examine how meaning is negotiated, contested, and distorted during active engagement and secondary circulation.

\subsection{Framing of ``Fact-Checking''}
We adopt the established framing of ``fact-checking,'' but this term has limitations when applied to data narratives. While fact-checking traditionally emphasizes the corroboration of discrete claims or events, data narratives also involve analytical premises, design choices, uncertainty, and rhetorical framing. Although modern fact-checking practices often address these interpretive dimensions, the term itself can imply categorical judgments that are not always appropriate for narratives involving ambiguity or partial validity. Future work should complement fact-checking perspectives with evaluative frames that foreground evidentiary warrant, degrees of confidence, and uncertainty.

\section{Conclusion}
We introduced TIC, a six-dimensional, process-oriented taxonomy of issues in data communication, grounded in prior scholarship and refined through qualitative analysis of 700 real-world data narratives. By organizing breakdowns across data, analysis, visual encoding, textual framing, reasoning, and audience interpretation, TIC offers a structured lens for examining how quantitative evidence is transformed into claims, representations, and arguments, as well as where this process can fail. Situated within a broader data communication pipeline, TIC further clarifies when issues emerge, how they propagate across stages and modalities, which actors are involved, and where sociotechnical interventions may be most effective. Beyond classification, TIC provides a foundation for identifying recurring pitfalls and designing future tools that help authors, editors, fact-checkers, and audiences scrutinize, contextualize, and mitigate problematic data communication.

\bibliographystyle{IEEEtran}
\bibliography{sections/bibliography,sections/references_z}

\begin{IEEEbiography}[{\includegraphics[width=1in,height=1.25in,clip,keepaspectratio]{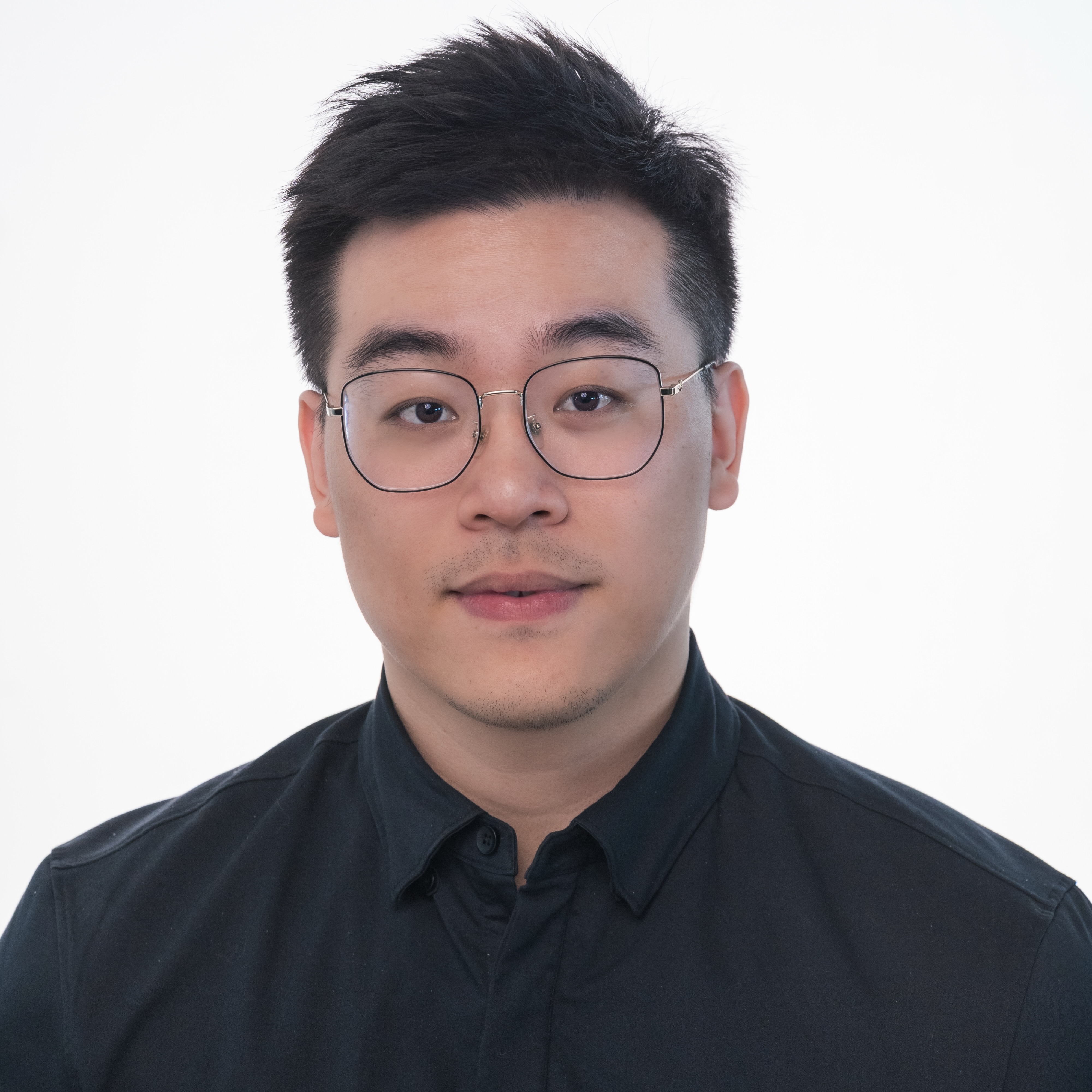}}]{Yu Fu}
is an Assistant Professor in Computer Science at the University of Central Florida. He received his Ph.D. in Human-Centered Computing from the Georgia Institute of Technology. His research lies at the intersection of data visualization and HCI, with an emphasis on designing interactive and intelligent systems that support data-driven communication. His work has been published in top-tier venues for visualization and human-computer interaction, including IEEE VIS, TVCG, CHI, and EuroVis.
\end{IEEEbiography}
\vspace{-4em}

\begin{IEEEbiography}[{\includegraphics[width=1in,height=1.25in,clip,keepaspectratio]{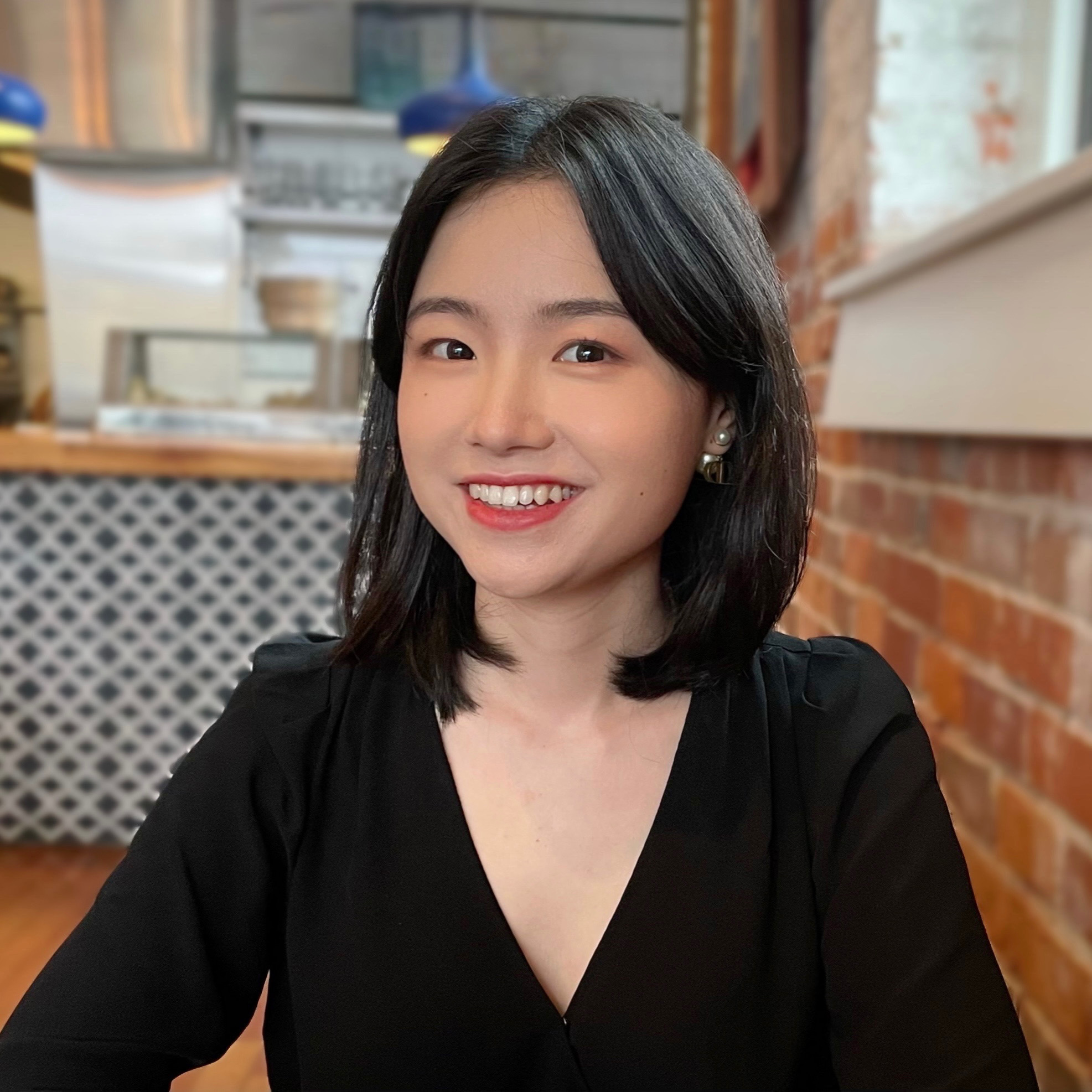}}]{Jiawei Zhou}
is a PhD candidate in Human-Centered Computing at Georgia Institute of Technology. Her research lies at the intersection of Human-AI Interaction, Social Computing, and Health \& Wellbeing, with an emphasis on AI safety and literacy and information quality. Her work has been published at top-tier venues for Human-Centered Computing such as CHI and CSCW with multiple paper awards.
\end{IEEEbiography}
\vspace{-4em}

\begin{IEEEbiography}[{\includegraphics[width=1in,height=1.25in,clip,keepaspectratio]{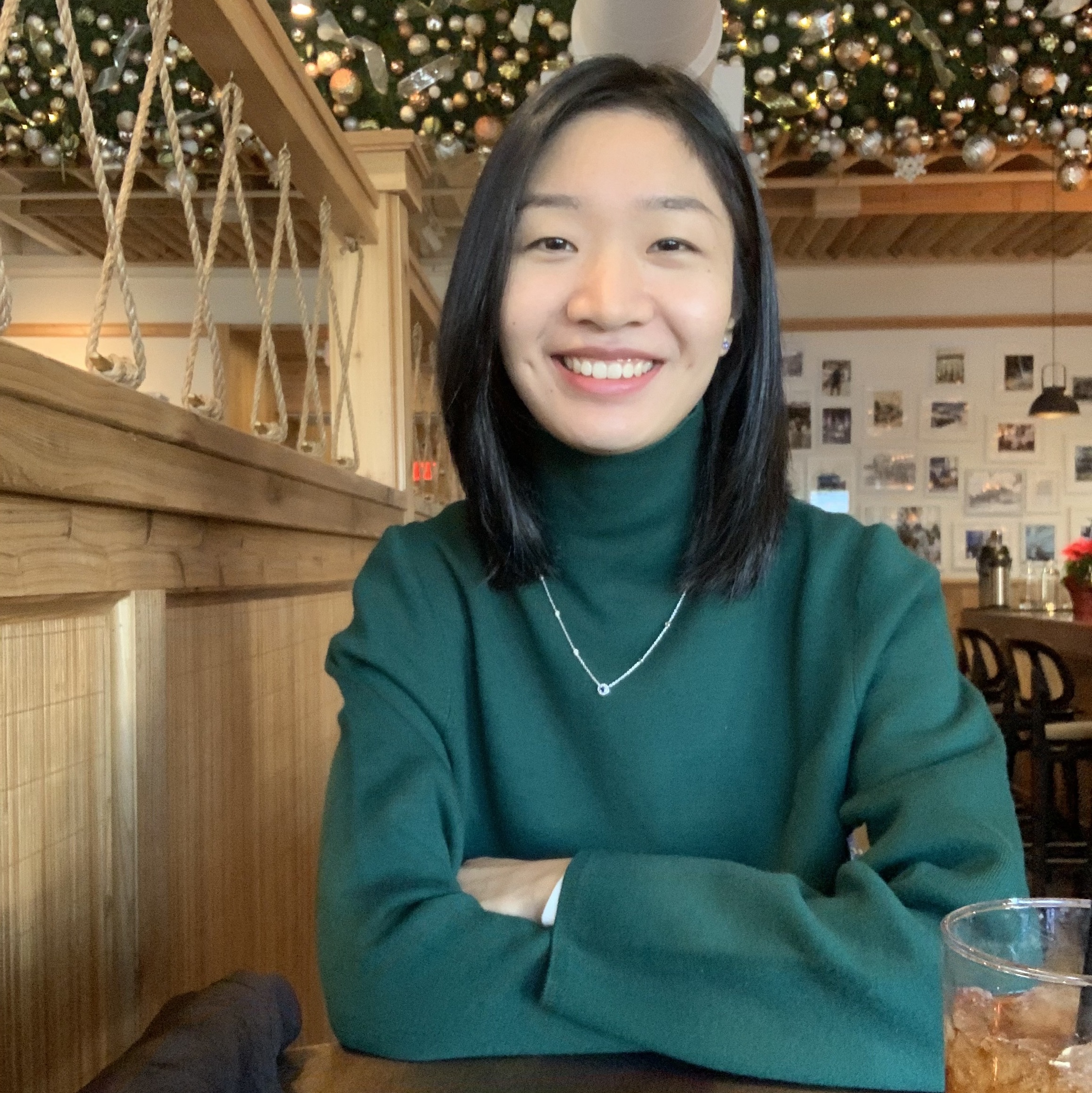}}]{Sichen Jin}
is a Ph.D. Candidate in Computer Science at the Georgia Institute of Technology. Her research focuses on data visualization, human-computer interaction, and geographic information science. She has published work on intelligent visual analytics systems for analyzing geospatial networks and high-dimensional spatiotemporal datasets in top-tier venues for GIS, visualization, and HCI.
\end{IEEEbiography}
\vspace{-4em}

\begin{IEEEbiography}[{\includegraphics[width=1in,height=1.25in,clip,keepaspectratio]{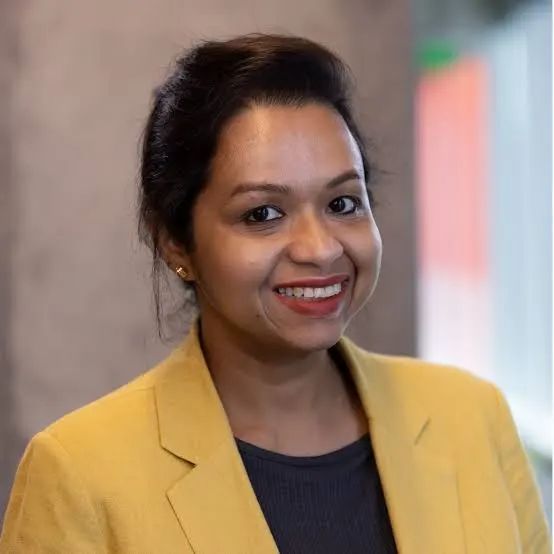}}]{Munmun De Choudhury}
is the J. Z. Liang Professor in the School of Interactive Computing at the Georgia Institute of Technology. Her research spans computational social science, human-computer interaction, and digital mental health, with widely recognized contributions to understanding well-being through social digital data. Her work has received major honors from SIGCHI, ICWSM, ACM-W, and the Web Science Trust, and has informed policy, advocacy, and public discourse on digital well-being.
\end{IEEEbiography}
\vspace{-4em}

\begin{IEEEbiography}[{\includegraphics[width=1in,height=1.25in,clip,keepaspectratio]{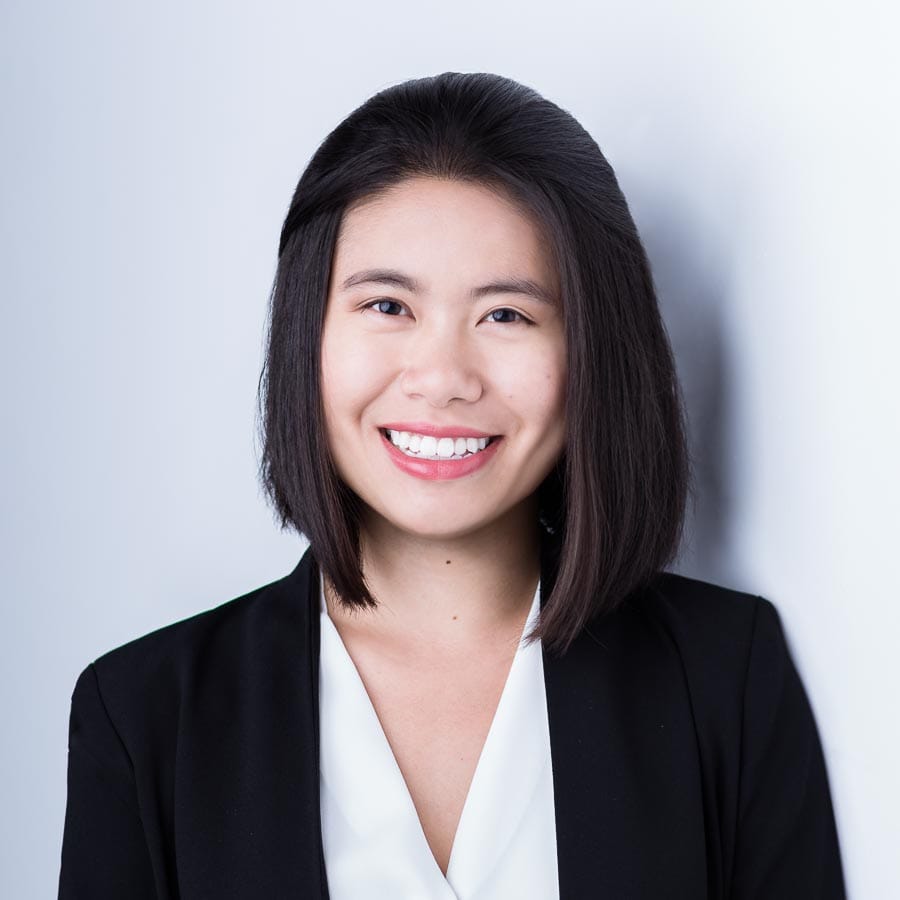}}]{Cindy Xiong Bearfield}
is an Assistant Professor in the School of Interactive Computing at the Georgia Institute of Technology. She designs visualizations that help people build calibrated trust in complex information, supporting thoughtful engagement with AI. Her work has been recognized with an NSF CAREER Award, a Google Research Scholar Award, and awards at leading HCI and visualization venues.
\end{IEEEbiography}
\vspace{-4em}

\begin{IEEEbiography}[{\includegraphics[width=1in,height=1.25in,clip,keepaspectratio]{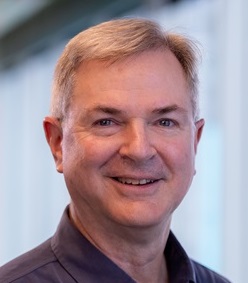}}]{John Stasko}
is a Professor Emeritus in the School of Interactive Computing at the Georgia Institute of Technology, where he previously directed the Information Interfaces Research Group. His research is in the areas of information visualization and visual analytics, approaching each from a human-computer interaction perspective. He was named an IEEE Fellow in 2014 and an ACM Fellow in 2022.
\end{IEEEbiography}

\end{document}